\documentclass{llncs}

\usepackage{latexsym}

\newcommand{\even} {\ensuremath{\mathcal{E}}}
\newcommand{\odd} {\ensuremath{\mathcal{O}}}
\newcommand{\Res} {\mbox{\tt Cl}}
\newcommand{\Rs} {\mbox{\tt Cls}}
\newcommand{\false} {\mbox{\tt  false}}

\newcommand{\vect}[1]{\ensuremath{\overline{#1}}}

\begin{document}

\title{ Detecting Unsolvable Queries for\\ Definite Logic Programs}
\titlerunning{Detecting Unsolvable Queries}
\author{ Maurice Bruynooghe\inst{1} \and Henk Vandecasteele\inst{1} \and
  D. Andre de Waal\inst{2} \and Marc Denecker\inst{1}}
\authorrunning{Maurice Bruynooghe et al.}
\tocauthor{Maurice Bruynooghe, Henk Vandecasteele, Marc Denecker
  (Katholieke Universiteit Leuven),
D. Andre de Waal (Potchefstroom University)
}
\institute{
Departement Computerwetenschappen\\
Katholieke Universiteit Leuven, Belgium \\
\email{\{maurice, henkv, marcd\}}@cs.kuleuven.ac.be,\\
\and Centre for Business
Mathematics and Informatics\\ 
Potchefstroom University for Christian Higher Education, South Africa \\
\email{BWIDADW}@puknet.puk.ac.za }

\maketitle

\begin{abstract}
  In solving a query, the SLD proof procedure for definite programs
  sometimes searches an infinite space for a non existing solution.
  For example, querying a planner for an unreachable goal state. Such
  programs motivate the development of methods to prove the absence of
  a solution. Considering the definite program and the query {\tt
    $\leftarrow$ Q} as clauses of a first order theory, one can apply
  model generators which search for a finite interpretation in which
  the program clauses as well as the clause {\tt false $\leftarrow$ Q}
  are true. This paper develops a new approach which exploits the fact
  that all clauses are definite. It is based on a goal directed
  abductive search in the space of finite pre-interpretations for a
  pre-interpretation such that $Q$ is false in the least model of the
  program based on it. Several methods for efficiently searching the
  space of pre-interpretations are presented.  Experimental results
  confirm that our approach find solutions with less search than with
  the use of a first order model generator.

\end{abstract}


\section{Introduction}
\label{sec:intro}

For many definite programs there exist queries for which the SLD-tree
(under the left-to-right computation rule) is infinite. In some cases
the infinite tree contains no solution. This paper is about proving the
latter efficiently. Our original motivation stems from the world of
planning. Typically, a planner searches an infinite space of candidate
plans for a plan satisfying all requirements. It is useful to have a
second process which searches for a proof that not all requirements
can be met; if found, the first process can be stopped (and the other
way around if a plan is found). Another application is in proving that
a program satisfies certain integrity constraints. For example, a
program defining even and odd numbers should satisfy the integrity
constraint that no number is both even and odd. This can be proven by
showing that the query {\tt $\leftarrow$ even(X), odd(X)} fails.

Failure of a query {\tt $\leftarrow$ Q} can be proven by showing that
{\tt Q} is not a logical consequence of the program, in other words by
constructing an interpretation in which the program clauses are true
(i.e.\ which is a model of the program) and in which {\tt Q} is false
(or alternatively, the clause {\tt false $\leftarrow$ Q} is true).
First order model generators \cite{FINDER1,FINDER,SEM,Peltier98} can
be used for this task. They search for an interpretation over a finite
domain such that all clauses of a given set evaluate to true.

This paper develops an alternative approach which exploits the fact
that definite programs have least models. If {\tt Q} is false in some
model based on a pre-interpretation, then it is also false in the
least model based on that pre-interpretation. Hence, it should be
better to search in the space of pre-int\-er\-pre\-ta\-tions for a
pre-interpretation such that {\tt Q} is false in the least model based
on it than to search in the larger space of interpretations for an
interpretation such that the program clauses are true and {\tt Q} is
false.

While this paper is one of the first to address the problem of proving
that an SLD-tree contains no solutions, the related problem, proving
that the SLD-tree is finite, has received a lot of attention in the
literature on termination analysis. See \cite{survey} for a survey. As
argued above, useful programs exist for which the SLD-tree is not
finite.

The problems caused by infinite branches in proof trees have also been
addressed in work on loop checking \cite{apt:iclp89,bol:tcs,ydshen}.
The idea is to monitor the execution and to try to prune infinitely
failing branches. However, methods have to choose between pruning too
much (causing incompleteness of the proof procedure) and preserving
completeness but missing some infinite branches. Proof procedures with
tabling such as XSB \cite{XSB:ACM} are perhaps a better alternative to
a completeness preserving procedure equipped with loop checking. They
avoid the overhead of monitoring the execution while using a different
proof procedure which, compared to SLD, reduces the number of infinite
branches. In particular, they always terminate for DATALOG programs.

The work in \cite{JLP92} takes a different approach. A logic program
is represented as a set of equivalence preserving rewrite rules. One
of the effects of the ``simplification inference rule'' is to
eliminate certain infinite derivations. We have not studied this
approach in detail. We suspect its power is comparable to that of
conjunctive partial deduction \cite{LeuschelDeSchreyeDeWaal:JICSLP96,CPD}
which is discussed in Section \ref{sec:alt}.

Section \ref{sec:prelim} recalls the basics about pre-interpretations
and introduces a trivial example. Section \ref{sec:prpr}, explains how
a pre-interpretation can be described by a number of facts, how a
program can be abstracted as a DATALOG program, and how the least
model based on that pre-interpretation can be queried by evaluating
the abstracted query on the DATALOG program. Section \ref{sec:search}
develops two procedures for proving failure. The first one combines
abduction with tabulation and failure analysis. The second one
combines constraints with tabulation and develops two alternatives for
the constraint solving. The first alternative uses abduction and failure
analysis in the constraint checking. The second one translates the
constraints into finite domain constraints and uses a finite domain
solver for the constraint checking. Section \ref{sec:alt} discusses
alternative approaches: the use of model generators for first order
logic \cite{FINDER1,FINDER,SEM,Peltier98}, of type analysis
\cite{Gal94,CharatonikPodelski-SAS98} (a query fails if its inferred
type is empty) and of program specialisation
\cite{LeuschelDeSchreyeDeWaal:JICSLP96,CPD} (the query fails if the
program ---for the given query--- can be specialised into the empty
program).  In section \ref{sec:exp}, the different approaches are
compared. Finally, in section \ref{sec:conc}, we draw some
conclusions. We assume some familiarity with the basics of tabulation,
e.g.\ \cite{OLDT,XSB:ACM,Memoing@CACM-92}.

Some of the authors of the current paper participated in a preliminary
investigation of the topic \cite{IJCAI97}. The current paper is an
extension of the work described in \cite{PLILP98}.

\section{Preliminaries}
\label{sec:prelim}

A pre-interpretation $J$ of a program $P$ consists of a domain $D =
\{d_1,\ldots,d_m \}$ and, for every functor $f/n$ a mapping
$f_J$ from $D^n$ to $D$. An interpretation $I$ based on a
pre-interpretation $J$ extends $J$ with a mapping $p_I$ from $D^n$ to
$\{{\mathit true,false}\}$ for every predicate $p/n$ in $P$. Extending
the language of the program $P$ with the domain $D$, an interpretation
can be identified with the set of atoms
$p(d_1,\ldots,d_n)$ for which $p_I(d_1,\ldots,d_n)$ is mapped to
${\mathit true}$. 

An interpretation $I$ is a model of a program $P$ iff all clauses of
$P$ are true under the interpretation $I$. For every
pre-interpretation $J$, a definite program has a model $I$ based on
$J$ (map $p_I(d_1,\ldots,d_n)$ to ${\mathit true}$ for all predicates
and all domain elements).  The intersection of a set of models based
on $J$ is also a model based on $J$, hence there is a unique least
model based on $J$. We denote this model by $LM_J(P)$. As a consequence, if an existentially quantified
conjunction $\exists{ \vect{X}} L_1 \wedge \ldots \wedge L_n$ (a
query) is false in a model based on a pre-interpretation $J$ then it
is also false in $LM_J(P)$.
So, to check whether some pre-interpretation $J$ can be the basis of a
model in which an existentially quantified conjunction is false, it
suffices to evaluate the conjunction in $LM_J(P)$. This can be
summarised in the following proposition:

\begin{proposition} \label{prop:least-model}
Given a pre-interpretation $J$, there exists an interpretation I based
on $J$  which is a model of $P$ $\cup$ \{{\tt false $\leftarrow$ Q}\} iff
$LM_J(P) \models$ {\tt false $\leftarrow$ Q}.
\end{proposition}

\begin{example} \label{ex:evenodd} Even/odd\\
  {\tt even(0) $\leftarrow$\\ even(s(X)) $\leftarrow$ odd(X).\\ 
    odd(s(X)) $\leftarrow$ even(X).\\
    }
  $D = \{\even,\odd\}$\\
  $0_J = \even$, $p_s(\even) = \odd$,
  $p_s(\odd) = \even$\\
  The least model is $\{even(\even), odd(\odd)\}$.  The query
  $\leftarrow$ {\tt even(X), odd(X)} fails because $\exists X even(X) \wedge
  odd(X)$ is false in this model. Executing the program with SLD or
  with a tabulating procedure (e.g.\ XSB  \cite{XSB:ACM})
  results in infinite failure. All methods discussed in Section
  \ref{sec:alt} can handle this problem.
\end{example}

A variable assignment $\sigma$ is a mapping from variables to domain
elements. Given a pre-interpretation $J$, this mapping can be extended
in a term assignment (variables are assigned according to $\sigma$,
functors according to $J$). $J_\sigma(t)$ denotes the term assignment
of $t$ under the pre-interpretation $J$ and variable assignment
$\sigma$. Given an interpretation $I$, the mapping can be further
extended to a truth assignment. $I_\sigma(F)$ gives the truth value of
the formula $F$ under the interpretation $I$ (based on a
pre-interpretation $J$) and the variable assignment $\sigma$.

\section{Proof procedures}
\label{sec:prpr}

In \cite{CD95}, Codish and Demoen developed so called abstract
compilation to perform groundness analysis of logic programs. Applying
a transformation on the original program, they obtain an abstracted
program. Groundness is then derived from a least model of the
abstracted program. In later work \cite{SAS94}, they used the same
technique to perform other analyses.  Boulanger and coauthors
\cite{Dima1,Dima2} explored the use of pre-interpretations to
approximate the s-semantics \cite{s-semantics} of programs and to
derive properties from this approximation.  In \cite{Gal}, the
approach of Codish and Demoen is generalised and presented as defining
a pre-interpretation and computing the least model of the abstracted
program; several applications are presented. It became clear that
abstract compilation is a technique allowing the efficient computation
of a program's least model based on a pre-interpretation.

Abstract compilation consists of eliminating non variable terms from
clauses. A term {\tt f(t1,\dots,tn)} is replaced by a fresh variable
{\tt X} and a call {\tt p$_f$(t1,\ldots,tn,X)} is added to the body of
the clause. This is repeated until all non variable terms have
disappeared from the program clauses. Note that this transformation is
independent from the particular pre-interpretation. The abstract
program is completed with the pre-interpretation $J$ in relational form:
the set of facts \{{\tt p$_f$(d$_1$,\ldots,d$_n$,d)$\leftarrow$} $|
f_J(d_1,\ldots,d_n) = d$\}. Each fact represents a {\em component} of
the pre-interpretation. With $J$ a pre-interpretation, $P^a_J$ denotes its
relational form.

\begin{example}
  Applying abstract compilation on the program and pre-interpretation
  of  Example \ref{ex:evenodd}, we obtain:\\
  {\tt
    even(X) $\leftarrow$ p$_0$(X)\\
    even(Y) $\leftarrow$ p$_s$(X,Y), odd(X).\\
    odd(Y) $\leftarrow$ p$_s$(X,Y), even(X).\\
    p$_0$(\even) $\leftarrow$\\
    p$_s$(\even,\odd) $\leftarrow$\\
    p$_s$(\odd,\even) $\leftarrow$\\
    } The clauses together with the facts of the pre-interpretation
  form a DATALOG program. The least model is $\{p_0(\even),
  p_s(\even,\odd), p_s(\odd,\even), even(\even), odd(\odd)\}$. The
  formula $\exists X even(X) \wedge odd(X)$ is false in this model. While
  the query $\leftarrow$ {\tt even(X), odd(X)} is nonterminating under
  SLD, it fails finitely under well known proof procedures such as
  bottom-up evaluation after magic-set transformation or top-down
  methods enriched with tabulation such as OLDT \cite{OLDT} and XSB
  \cite{XSB:ACM}.
\end{example}

In what follows, we give some results formalising the relationship
between a program and its abstraction. First, we introduce some
notational conventions. With $Cl$ a clause, $Cl^a$ denotes its
abstraction; with $P$ a set of clauses (a program), $P^a$ denotes its
abstraction.  $J^a$ is the Herbrand pre-interpretation of $P^a \cup
P^a_J$. Remark that $J^a$ has the same domain as $J$ as the domain
elements are the only functors in $P^a \cup P^a_J$.  $I_J$ denotes the
interpretation which is the least Herbrand model of $P^a_J$. Finally,
$I^a$ is the interpretation of $P^a \cup P^a_J$ corresponding to the
interpretation $I$ when $I^a = I \cup I_J$.

\begin{theorem}\label{th:equiv}
  Let $I$ be an interpretation of $P$ based on pre-interpretation $J$
  and $I^a$ the corresponding interpretation of $P^a \cup P^a_J$.  The
  interpretation $I$ is a model of $P$ $\cup$ \{{\tt false
    $\leftarrow$ Q}\} iff $I^a$ is
  a model of $P^a \cup P^a_J$ $\cup$ \{{\tt false $\leftarrow$ Q$^a$}\}.\\
  {\em Proof.

    Consider a slightly different clause transformation which replaces
    a term {\tt f(t1,\dots,tn)} by a fresh variable {\tt X} and adds
    the equality {\tt f(t1,\dots,tn) = X} to the body of the clause.
    Repeat this transformation until all non variable terms are
    eliminated from argument positions in program predicates and from
    argument positions in terms of the equalities. The difference
    between the resulting clause $Cl'$ and the abstracted clause
    $Cl^a$ is that $Cl'$ has an equality {\tt f(X1,\ldots,Xn) = X}
    where $Cl^a$ has a call {\tt p$_f$(X1,\ldots,Xn,X)}.

    This new transformation is equivalency preserving hence $I \models
    Cl$ iff $I \models Cl'$.
    
    Given a variable assignment $\sigma$, $I_\sigma(f(X1,\ldots,Xn) =
    X)$ is true iff $\sigma(X) = f_J(\sigma(X1),\ldots,\sigma(Xn))$.
    We also have that $I^a_\sigma(p_f(X1,\ldots,Xn,X))$ is true iff
    $\sigma(X) = f_J(\sigma(X1),\ldots,\sigma(Xn))$. Hence,
    $I_\sigma(f(X1,\ldots,Xn) = X))= I^a_\sigma(p_f(X1,\ldots,Xn,X))$.
    Also for program predicates, we have $I_\sigma(p(X1,\ldots,Xn))=
    I^a_\sigma(p(X1,\ldots,Xn))$, hence $I \models Cl'$ iff $I^a
    \models Cl^a$. As this holds for all clauses and for the query,
    the theorem follows.  \qed }
\end{theorem}

Abstract compilation is a simple variant of more general
transformation to substitute predicate symbols by function symbols
which is well known in logic, e.g.\ \cite{Enderton72}.

\begin{corollary}
Given a sound and complete proof procedure, the query {\tt
  $\leftarrow$ Q$^a$} fails for the program $P^a \cup P^a_J$ iff there
  exists an interpretation $I$ based on $J$ such that $I(Q)$ is
  false.\\
{\em Proof\\
By Prop. \ref{prop:least-model}, $I(Q)$ is false iff $LM_J(P) \models$
{\tt false $\leftarrow$ Q}.\\
By Theorem \ref{th:equiv}, $LM_J(P) \models$ {\tt false $\leftarrow$
  Q} iff $LM_J(P^a) \cup I_J \models$ {\tt false $\leftarrow$ Q$^a$}.\\
$LM_J(P^a) \cup I_J$ is the least Herbrand model of $P^a \cup P^a_J$,
hence, given a sound and complete proof procedure, {\tt $\leftarrow$
  Q$^a$} fails iff $LM_J(P^a) \cup I_J \models$ {\tt false $\leftarrow$
  Q$^a$} iff $I(Q)$ is false. \qed}
\end{corollary}

\section{The search for the right pre-interpretation}
\label{sec:search}

To prove failure, the approach is to select a domain and a
pre-interpretation and to show finite failure when executing the
abstracted query with the abstracted program. A straightforward way
consists of selecting a domain and trying all pre-interpretations
until one is found for which the query fails. If none exists, one can
try again with a larger domain. However, for programs with a
substantial number of function symbols and constants, this quickly
results in a very large search space. Indeed, with a $n$-element
domain, an $m$-ary functor has $n^{(n^m)}$ possible
pre-interpretations.

Hence better is to consider $P^a_J$, the part defining the
pre-interpretation, as unknown and the use a procedure which can guess
the missing predicate definitions. Abduction \cite{Kakas} is such a
mechanism. In an abductive setting, given is a logic program $P$
defining a subset $D$ of its predicates, a set $T$ of integrity
constraints in classical logic and some query $Q$. Abduction searches
a definition $\Delta$ of the open, abducible predicates, i.e. those
not defined by $P$ such that $P\cup \Delta \models Q$ and $P \cup
\Delta \models T$. For our problem, the defined predicates are those
defined by $P^a$ (the set of abstracted clauses), the abducibles are
$p_f/n+1$, the query is the abstracted query {\tt $\leftarrow$ not
  Q$^a$} (we want a solution for which the query fails) and the
integrity constraints are axioms restricting $p_f/n+1$ to correspond
to a pre-interpretation which is a total function, i.e.\ $\forall
X_1\ldots X_n \exists! Y. p_f(X_1,\ldots,X_n,Y)$. Hence the problem is
to find a $\Delta$ such that $P^a \cup \Delta \models$ {\tt false
  $\leftarrow$ Q} and satisfies the integrity constraints.

In a first experiment, we have used the general purpose abductive
procedure SLDNFA \cite{SLDNFA} to solve the problem. A first problem
that we met was that SLDNFA looped on the abstract program due to
non-acyclic recursion in it. To overcome this
problem, the clauses defining the recursive predicates were
transformed into integrity constraints.  Some initial experiments
showed the feasibility, but also the need for a dedicated procedure
which allows to experiment with different control strategies.

Abduction is complex in the general case due to the presence of
variables in abductive calls and the interaction with negation as
failure. We are only concerned with definite programs, that simplifies
substantially the design of a dedicated procedure. Moreover, we know
that the pre-interpretation of a functor $f/n$ is a total function
from $D^n$ to $D$. Hence, a complete pre-interpretation has exactly
one fact {\tt p$_f$(d1,\ldots,dn,d) $\leftarrow$} for every tuple
$\langle d1,\ldots,dn \rangle$ of domain elements and $d$ is also a
domain element. As a consequence, our abductive procedure has only to
``guess'' among the domain elements for the value of $d$. So it
becomes fairly simple to ensure correctness and completeness. A call
{\tt p$_f$(t1,\ldots,tn,t)} needs to be resolved with all the facts
{\tt p$_f$(d1,\ldots,dn,d)} such that $\langle d1,\ldots,dn \rangle$
unifies with $\langle t1,\ldots,tn \rangle$. If some of these facts do
not yet exist, they have to be abduced. To obtain an exhaustive search
over all candidate abductive solutions, one has only to take care that
all domain elements are in turn considered as candidate values for
$d$.

To overcome the problem of looping, our dedicated procedure makes use
of tabulation. Tabulation \cite{OLDT,XSB:ACM,Memoing@CACM-92} avoids
non termination in the case of DATALOG programs. As we only consider
definite programs, the concept is fairly simple. When a call {\tt
  p(t1,\ldots,tn)} to a tabled predicate is selected in a goal, the
goal is suspended and the query {\tt $\leftarrow$ p(t1,\ldots,tn)} is
evaluated in isolation of its goal. Eventually, evaluation of this new
query leads to computed answer substitutions $\sigma_1, \sigma_2,
\ldots$. For each of these answers $\sigma_i$, an answer lemma {\tt
  p(t1,\ldots,tn)$\sigma_i \leftarrow$} is stored (if not the renaming
of a previous answer) in the table associated with the call {\tt
  p(t1,\ldots,tn)}. The suspended goal is reactivated for each answer
lemma and a resolution step is performed, using the answer lemma as
program clause. If in another goal the atom {\tt p(s1,\ldots,sn)} is
selected and the atom happens to be a renaming of {\tt
  p(t1,\ldots,tn)}, then no separate query for {\tt $\leftarrow$
  p(s1,\ldots,sn)} is launched. The goal is simply suspended and is
reactivated for each answer lemma stored in the table associated with
the query {\tt $\leftarrow$ p(t1,\ldots,tn)}.

As the abstract program has no functors apart from the 0-arity domain
elements, only a finite number of distinct calls can occur. Also, for
each call, only a finite number of distinct answers can occur, hence
termination is ensured.

We prefer top-down evaluation with tabulation above bottom-up
evaluation because top-down is goal directed. Our procedures use
heuristics; these try to find those refutations which are short and use
as few different components of the pre-interpretation as possible
first. We find it more convenient to design such heuristics in the
context of a top-down procedure.

As a final remark, it is well possible that not all components of a
pre-interpretation are needed to evaluate a particular abstract query.
In such case, our dedicated procedure will not abduce the complete
pre-interpretation. For the not abduced facts {\tt
  p$_f$(d1,\ldots,dn,d)$\leftarrow$}, any value $d$ can be chosen.
Adding them will neither modify the proof structure nor the outcome of
the query evaluation.

\subsection{An abductive approach}
\label{sec:abductive}

When experimenting with a dedicated abductive procedure, an early
observation concerned the tabulation mechanism. Typically, several
table entries were created for the same predicate. Frequently, a
``final'' call occurred with a call pattern where all arguments are
free variables. This final call subsumed all previous ones. Hence it
is more efficient, when a call {\tt p(t1,\ldots,tn)} occurs to a
tabled predicate, to compute once and for all the answers to the most
general query {\tt $\leftarrow$ p(X1,\ldots,Xn)}. Unification of the
call {\tt p(t1,\ldots,tn)} with the answer lemmas then selects the
answers to the call which occurred. We also observed that it was
preferable to table all program predicates, whether recursive or not.
Again the sum of the costs of evaluating each separate call was larger
than evaluating once the most general call. We adopted these
strategies in all our systematic experiments and it is hard wired in
the description of the procedure below.

We need some notational conventions to describe the procedure. The
state of the computation is represented as a set of clauses. The
symbol \Res\ represents a clause and the symbol \Rs\ represents a set
of clauses.  Clauses that are a renaming of each other are considered
equal.  {\tt As} and {\tt Bs} represent sequences of atoms. A clause
is represented as {\tt H $\leftarrow$ As} in which the head {\tt H} is
an atom (or {\tt false}). {\tt s} and {\tt t} ({\tt \vect{s}} and {\tt
  \vect{t}}) denote a term (a vector of terms); {\tt d} ({\tt \vect{d}})
denotes a domain element (a vector of domain elements).

The procedure (Fig.\ \ref{fig:abd}) is described as a set of inference
rules which extend the state with new clauses.  Given a query {\tt
  $\leftarrow$ As}, the initial state of the derivation is represented
as \{{\tt \false\ $\leftarrow$ As}\}. {\tt p/n} refers to a predicate
of the original program; calls to such predicates are tabled. The
tabulation has an implicit representation: a predicate is tabled when
the clauses defining it occur in the state; answer lemmas for {\tt
  p/n} occur in the state as facts {\tt p(\vect{s})$\leftarrow$}
(when all calls in the body of a clause defining {\tt p/n} are solved,
a fact is left).  {\tt abduce$_f$(\vect{t}) } is the notation we use
for a call to an abducible predicate {\tt p$_f$(\vect{t})} of the
pre-interpretation.  These calls are not tabled. A clause {\tt H
  $\leftarrow$ Lookup(p(\vect{t})),As} is a suspended clause, waiting
for answers of the call {\tt p(\vect{t})}.  For simplicity of
representation we assume the computation rule always select the
leftmost atom in the body of a clause.

\begin{figure}[htbp]
  \begin{center}
    \leavevmode

\begin{tabular}{||l|l|l|l||}
\hline 
Nr& \multicolumn{1}{c|} {State}& \multicolumn{1}{c|}{Condition} &
\multicolumn{1}{c||} {New State} \\ 
\hline\hline
1a&\{{\tt H $\leftarrow$ p(\vect{t}),As}\} & not\_tabled({\tt p}) & \{{\tt
  H $\leftarrow$ Lookup(p(\vect{t})),As}\} \\
& $\cup$ {\tt \Rs} &  & $\cup$ {\tt \{\Res $|$ \Res} is a clause defining {\tt
  p}\}\\
& & & $\cup$ {\tt \Rs}\\
\hline
1b& \{{\tt H $\leftarrow$ p(\vect{t}),As}\}& tabled({\tt p}) & \{{\tt
  H $\leftarrow$ Lookup(p(\vect{t})),As}\} \\
& $\cup$ \Rs &  &$\cup$ {\tt \Rs}\\
\hline
2& {\tt \Rs} which contains  & unify({\tt\vect{s}},{\tt \vect{t}})&  
\{{\tt (H $\leftarrow$ As)mgu(\vect{t},\vect{s})}\}  \\
& {\tt p(\vect{s})$\leftarrow$}  and  & &$\cup$ {\tt \Rs} \\
& {\tt H $\leftarrow$ Lookup(p(\vect{t})),As}& & \\
\hline
3& {\tt \Rs} which contains & unify({\tt\vect{d}},{\tt \vect{t}})&
\{{\tt (H $\leftarrow$ As)mgu(\vect{t},\vect{d})}\}  \\
&{\tt abduce$_f$(\vect{d})$\leftarrow$ } and & &$\cup$ {\tt \Rs} \\
&{\tt H $\leftarrow$ abduce$_f$(\vect{t}),As}& & \\
\hline
4&  {\tt \Rs} which contains & $\exists$ \vect{d}:
(unify(\vect{t},\vect{d}) and  $\forall$ d:&\{{\tt
  abduce$_f$(\vect{d},$d$)$\leftarrow$}\}$\cup$ {\tt \Rs}  \\  
&{\tt H $\leftarrow$ abduce$_f$(\vect{t},$t$),As}&
{\tt abduce$_f$(\vect{d},$d$)$\leftarrow$} $\notin$  {\tt \Rs})
&  where $d$ is a domain element\\
\hline
5&  {\tt \Rs} &{\tt false$\leftarrow$} $\in$ {\tt \Rs}  & failure\\
\hline
\end{tabular}
    \caption{Inference rules of abductive procedure.}
    \label{fig:abd}
  \end{center}
\end{figure}

We assume a fixed number of domain elements. Rules 1 to 3 describe the
execution under tabling. Given a complete pre-interpretation, these
rules will derive all consequences. Rule 1 handles a call to a tabled
predicate. The clauses defining it are added to the state when it is
the first call to the predicate (1a). Whether the first call or not,
the clause is suspended (the selected call is wrapped inside {\tt
  Lookup}). Rule 2 uses an answer lemma to derive a new clause from a
suspended one and rule 3 uses a fact from the pre-interpretation to
derive a new clause (here and elsewhere, the necessary renaming is
omitted to simplify the presentation). Rule 4 performs abduction on
demand: if rule 3 needs a component of the pre-interpretation which is
not yet defined, then rule 4 adds such a component to the state.  The
value assigned to the component is chosen from the domain. This rule
is non-deterministic; it derives one new state for each domain
element. Rule 5 detects that the query has a solution, i.e.\ that the
(partial) pre-interpretation being considered does not meet the
requirement. Hence it replaces the state by the final state {\tt
  failure}.

A solution is found when a final state is reached that is different
from failure, (no new clauses can be derived). Although the
pre-interpretation can still be partial, any extension to a complete
pre-interpretation is the basis for a model of the program in which
the query fails.

\paragraph{Correctness and termination.}

Assuming the pre-interpretation (the set of facts {\tt
  abduce$_f$(\vect{d})}) is complete, rules 2 and 3 perform unit
resolution on a set of clauses consisting of the query and the program
clauses. Rule 1 is a heuristic which delays resolution steps on
program clauses until it is certain they can contribute in the
derivation of the empty clause from the query. Unit resolution is
known to be a correct and complete proof procedure for definite
clauses; moreover, as the Herbrand universe of the program is the
finite domain of the pre-interpretation, only a finite number of
clauses can be derived and termination is ensured (for a fixed
pre-interpretation).
  
However, the computation start with an empty pre-interpretation. The
role of rule 4 is to create a component of the pre-interpretation as
soon as rule 3 can make use of it. This does not affect correctness.
As rule 4 can only choose among a finite number of domain elements,
and can only be applied a finite number of times in a particular
derivation, overall termination remains ensured. Rule 5 plays the role
of a filter, it stops the derivation when it detects that the query
has a solution.

\paragraph{Control.}

In our implementation, the choice offered by rule 4 is implemented
through enumeration and backtracking. When
applying rule 4, a choice point is created which keeps track of the
untried domain elements. When rule 5 is activated, it triggers
backtracking to the last choice point where the next value is
tried. If none, backtracking continues to the previous choice
point. The overall computation fails as soon as backtracking occurs
and no alternative is left in any choice point.

The order in which rules are applied can have a big impact on the size
of the search space. Obviously, rule 5, which triggers backtracking
should be activated as soon as a clause {\tt (false$\leftarrow$)} is
inferred. Also, the creation of a new choice point should be delayed
as long as possible, i.e.\ one should not create another choice point
if the already abduced facts allow to infer the clause {\tt
  (false$\leftarrow$)}. A simple implementation almost realising this 
strategy selects the leftmost literal in a clause body and applies the
applicable rule, delaying the processing of the clause if the selected
literal is an abducible for which not all instances are already
abduced. Once no other rules are applicable, rule 4 is applied on one
of the delayed clauses. Experiments showed that more complex
strategies ---causing more meta-interpretation overhead--- which do
not necessarily select the leftmost literal tend to perform better. In
processing new clauses, the system reported in section \ref{sec:exp}
has a preference for applying a look-up step (rule 2). If none is
applicable it attempts to apply rule 3 on an abducible literal for
which all matching components of the pre-interpretation are defined
(so that all resolvents can be computed at once and the original
clause can be removed). If the clause has no such abducible, it is
delayed. Once all clauses are delayed, it gives preference to the
application of rule 1 on a clause with only calls to program
predicates. If no such clause is delayed, rule 4 is applied on a
delayed clause. A heuristic selects the delayed clause containing the
abducible which needs the least number of yet to be abduced instances
to allow for the application of rule 3 and abduces those instances.

\subsection{A constraint approach}
\label{sec:constr}

We use a more compact notation in this section. We write {\tt
  abduce(f(a),X)} instead of {\tt abduce$_a$(Y), abduce$_f$(Y,X)} With
this notation, the abstraction of an atom {\tt p(f(a))} becomes {\tt
  abduce(f(a),X), p(X)}.

A weakness of the abductive approach can be illustrated with the
following example: assume that the pre-interpretation of a functor
{\tt f/1} has already been abduced as {\tt abduce(f(d1),d1)} and {\tt
  abduce(f(d2),d2)} and that a clause {\tt false $\leftarrow$
  abduce(f(g(h(a))),X), abduce(g(h(a)),X)} is derived. The
pre-interpretation of {\tt f/1} is such that for all domain elements
$d$, $f(d)=d$, hence, whatever the pre-interpretation for {\tt a/0, h/1},
and {\tt g/1}, {\tt false$\leftarrow$} will be derived. The abductive
system will abduce pre-interpretations for {\tt a/0, h/1}, and {\tt g/1},
and will then discover the failure. It will exhaustively enumerate all
pre-interpretations for {\tt a/0, h/1}, and {\tt g/1} before backtracking to
the pre-interpretation of {\tt f/1}.

A constraint based approach can to a large extent avoid such problems.
We consider the abducibles as constraints and use a special purpose
constraint solver which checks the existence of a pre-interpretation
which satisfies all constraints. In the above example, if the
pre-interpretation of {\tt f/1} is constrained to the shown one and
the clause {\tt false $\leftarrow$ abduce(f(g(h(a))),X),
  abduce(g(h(a)),X)} is derived, then the solver detects the
inconsistency and triggers backtracking.

This approach makes it necessary to reformulate our abductive
system. The major difference is with respect to the tabulation. The
answers to a tabled predicate are no more simple facts but
constrained facts (of the form {\tt p(\vect{X}) $\leftarrow$
  abduce(\ldots), \ldots, abduce(\ldots)}). A problem is that one can
have an infinite number of syntactically different answers. However,
with a finite domain and a fixed pre-interpretation, the set of
answers  (its model) is finite. So it must be
possible to add constraints which enforce the finiteness.
Before presenting the formal system, we illustrate
the main ideas with the {\tt even/odd} example. 

\begin{example} Even/odd\\
  The program is as follows:\\
  {\tt
    even(X) $\leftarrow$ abduce(0,X).\\
    even(Y) $\leftarrow$ abduce(s(X),Y), odd(X).\\
    odd(Y) $\leftarrow$  abduce(s(X),Y), even(X).
    }
  
  The execution is shown in Fig.\ \ref{fig:ex}.  We represent the
  state of the derivation by three components, the set of clauses, the
  set of answers and the constraint store which holds the set of
  constraints (as before, the component with the fixed abstract
  program is left out). $\emptyset$ stands for the empty set. {\em
    Lookup} is abbreviated as {\em L}, and {\em false, abduce, even}
  and {\em odd} respectively as {\em f, ab, e} and {\em o}. Finally,
  {\em sb} is the abbreviation of {\em subsumed}.

  \begin{figure}[htbp]
    \begin{center}
      \leavevmode

  \begin{array}[t]{|r|l|l|l|}
\hline
&\multicolumn{1}{c|} {\rm Clauses} & \multicolumn{1}{c|}{\rm Answers}
&\multicolumn{1}{c|} {\rm Constraint\ Store}\\ 
\hline\hline
0& f \leftarrow e(X), o(X)&\emptyset &\emptyset \\
\hline
1& f \leftarrow L(e(X)), o(X)&\emptyset &\emptyset \\
 & e(X) \leftarrow ab(0,X)& & \\
 & e(Y) \leftarrow ab(s(X),Y), o(X)& & \\
\hline
2& f \leftarrow L(e(X)), o(X)&e(X) \leftarrow ab(0,X)&\emptyset \\
 & e(Y) \leftarrow ab(s(X),Y), o(X)& & \\
\hline
4& f \leftarrow L(e(X)), o(X)&e(X) \leftarrow ab(0,X)&\emptyset \\
 & e(Y) \leftarrow ab(s(X),Y), L(o(X))& & \\
 & o(Y) \leftarrow  ab(s(X),Y), L(e(X))& & \\
\hline
6& f \leftarrow L(e(X)), o(X)& e(X) \leftarrow ab(0,X) &\emptyset \\
 & f \leftarrow ab(0,X), o(X)& & \\
 & e(Y) \leftarrow ab(s(X),Y), L(o(X))& & \\
 & o(Y) \leftarrow  ab(s(X),Y), L(e(X))& & \\
 & o(Y) \leftarrow  ab(s(0),Y)& & \\
\hline
8& f \leftarrow L(e(X)), o(X)& e(X) \leftarrow ab(0,X) & \emptyset\\
 & f \leftarrow ab(0,X), L(o(X))& o(Y) \leftarrow ab(s(0),Y) &\\  
 & e(Y) \leftarrow ab(s(X),Y), L(o(X))& &\\
 & o(Y) \leftarrow  ab(s(X),Y), L(e(X))& &\\
\hline
10& f \leftarrow L(e(X)), o(X)& e(X) \leftarrow ab(0,X) & \emptyset\\
 & f \leftarrow ab(0,X), L(o(X))& o(Y) \leftarrow ab(s(0),Y) &\\ 
 & f \leftarrow ab(0,X), ab(s(0),X)& & \\ 
 & e(Y) \leftarrow ab(s(X),Y), L(o(X))& &\\
 & e(Y) \leftarrow ab(s(s(0)),Y)& &\\
 & o(Y) \leftarrow  ab(s(X),Y), L(e(X))& &\\
\hline
11& f \leftarrow L(e(X)), o(X)& e(X) \leftarrow ab(0,X) & 
                                            f \leftarrow ab(0,X), ab(s(0), X)\\
 & f \leftarrow ab(0,X), L(o(X))& o(Y) \leftarrow ab(s(0),Y),& \\ 
 & e(Y) \leftarrow ab(s(X),Y), L(o(X))& & \\
 & e(Y) \leftarrow ab(s(s(0))),Y), & &\\
 & o(Y) \leftarrow  ab(s(X),Y), L(e(X))& &\\
\hline
12& f \leftarrow L(e(X)), o(X)& e(X) \leftarrow ab(0,X) & f \leftarrow
                       ab(0,X), ab(s(0),X)\\
 & f \leftarrow ab(0,X), L(o(X))& o(Y) \leftarrow ab(s(0),Y) &
                                        sb(e(Y) \leftarrow ab(s(s(0)),Y),\\ 
 & e(Y) \leftarrow ab(s(X),Y), L(o(X))& & 
                            \hspace{1.5em}   \{  e(X) \leftarrow ab(0,X)\})\\
 & o(Y) \leftarrow  ab(s(X),Y), L(e(X))& & \\ 
\hline
  \end{array}

      \caption{Constraint based execution for even-odd program.}
      \label{fig:ex}
    \end{center}
  \end{figure}

\begin{description}
\item[0.] In the initial state the only clause is the query.  The
  leftmost atom is selected; the two clauses defining {\tt even/1}
  are added to the set of clauses and the original clause suspends,
  waiting for answers from {\tt even/1}.
\item[1.] The second clause, a constrained fact, is selected. A choice
  point is created.  The first alternative adds the constraint {\tt
    subsumed(even(X) $\leftarrow$ abduce(0,X),$\emptyset$)} to the
  constraint store in an attempt to have the new fact subsumed by the
  existing ones.  Whatever the pre-interpretation, {\tt abduce(0,d)}
  is true for some domain element $d$ and {\tt even(d)} is an answer
  which is not subsumed by previous answers as there are none. Hence
  the constraint is false and the second and last alternative is
  taken: the fact is added to the set of answers and the constraint
  {\tt not(subsumed(even(X) $\leftarrow$ abduce(0,X),$\emptyset$))} is
  added to the store. The constraint is equivalent to true, hence the
  store remains empty.
\item[2.] The call {\tt odd(X)} is selected in the second clause; the
  clause defining {\tt odd/1} is added. In the (omitted) new state,
  this clause is chosen and its atom {\tt even(X)} is selected. As
  {\tt even/1} has been called before, no new clauses are added.
\item[4.] This step and the next one perform resolution between the
  answer and suspended first and last clause. In one of the new clauses,
  {\tt abduce(s(X),Y), abduce(0,X)} is abbreviated as {\tt
    abduce(s(0),Y)}.
\item[6.] One step selects the atom {\tt odd(X)} in the second clause
  and suspends the clause. Another step selects the last clause which
  is a constrained answer. We have a choice point; as there are no
  previous answers for {\tt odd/1}, the subsumption constraint leads
  again to an inconsistent store and the not-subsumption constraint is
  again equivalent to true. Hence, the net effect is that the clause is
  added to the answers.
\item[8.] The next two steps perform resolution between the new answer
  and the second and third clause, resulting in two new clauses.
\item[10.] The third clause is a constraint, consistent with the store,
  and is added to it. The constraint says that {\tt 0} and {\tt s(0)}
  have to be different under the pre-interpretation.
\item[11.] The fourth clause is an answer for {\tt even/1}. A choice
  point is created. The first alternative creates the constraint that
  the new answer is subsumed by the existing answers: {\tt
    subsumed(even(Y) $\leftarrow$ abduce(s(s(0)),Y), \{even(X)
    $\leftarrow$ abduce(0,X)\})}. It is consistent with the constraint
  store (e.g.\ with $p_0 = \even$\ \ \ $p_s(\even) = \odd$\ \ \ 
  $p_s(\odd) = \even$, {\tt subsumed(even(\even), \{even(\even)\})} is
  true), hence it is added to it and the answer clause is dropped. The new
  constraint says that $0$ and $s(s(0))$ have to be equal under the
  pre-interpretation.
\item[12.] No new clauses can be derived. The store is consistent,
  hence there exists a pre-interpretation satisfying it (e.g.\ {\tt
    abduce(0,d1), abduce(s(d1),d2)} and {\tt abduce(s(d2),d1)}) and
  {\tt $\leftarrow$ even(X), odd(X)} is false in the least model based
  on a pre-interpretation consistent with the constraint store.
\end{description}

\end{example}

The inference rules of the constraint procedure are shown in Fig.\ 
\ref{fig:constr}.  A state, consisting of clauses, answers and a
constraint store is represented as {\tt \Rs} $\Diamond$ Answ
$\Diamond$ Store.
The symbols {\tt As} and {\tt Bs} stand for any sequence of
atoms, while {\tt Abds} stands for a sequence consisting solely of
abduce atoms.  {\tt Store} stands for a conjunction (set) of
constraints, {\tt Answ} for a set of answers (constrained facts) and
{\tt Answ$_p$} for the subset of answers about predicate {\tt p} . The
initial state is given by {\tt false $\leftarrow$ As $\Diamond$
  $\emptyset$ $\Diamond$ $\emptyset$} where {\tt $\leftarrow$ As} is
the query. subs($x,y$) is an abbreviation for subsumed($x,y$) and
not\_subs($x,y$) for not(subsumed($x,y$)); inconst($x$) is an
abbreviation for inconsistent($x$).  Remember that arguments of
program predicates of the abstracted program are always variables.

\begin{figure}[htbp]
  \begin{center}
    \leavevmode

\begin{tabular}[t]{||l|l|l|l||}
\hline 
Nr& \multicolumn{1}{c|} {State}& \multicolumn{1}{c|}{Condition} &
\multicolumn{1}{c||} {New State} \\ 
\hline\hline

1a&\{{\tt H $\leftarrow$ p(\vect{t}),As}\} $\cup$ \Rs &
not\_tabled({\tt p}) & \{{\tt  H $\leftarrow$
  Lookup(p(\vect{t})),As}\} $\cup$ {\tt \Rs} \\
&$\Diamond$ Answ $\Diamond$ Store  &  & $\cup$ {\tt \{\Res $|$
  \Res} is a clause defining {\tt p}\}\\ 
& & &  $\Diamond$  Answ $\Diamond$ Store\\
\hline

1b& \{{\tt H $\leftarrow$ p(\vect{t}),As}\} $\cup$ {\tt \Rs} &
tabled({\tt p}) & \{{\tt 
  H $\leftarrow$ Lookup(p(\vect{t})),As}\} $\cup$ {\tt \Rs} \\
&$\Diamond$  Answ $\Diamond$ Store &  & $\Diamond$  Answ $\Diamond$ Store\\
\hline

2& {\tt \Rs}  $\Diamond$  Answ $\Diamond$ Store  &
unify({\tt\vect{s}},{\tt \vect{t}})&   
\{{\tt (H $\leftarrow$ Abds, As)mgu(\vect{t},\vect{s})}\} $\cup$ {\tt
  \Rs} \\
& {\tt p(\vect{s}) $\leftarrow$ Abds} $\in$ Answ  & &  $\Diamond$  Answ
$\Diamond$ Store   \\
& {\tt H $\leftarrow$ Lookup(p(\vect{t})),As} $\in$  {\tt \Rs}& & \\
\hline

3& \{{\tt false $\leftarrow$ Abds}\} $\cup$ {\tt \Rs} & &  {\tt \Rs}
$\Diamond$  Answ \\ 
& $\Diamond$  Answ $\Diamond$ Store & & $\Diamond$  \{{\tt false
  $\leftarrow$ Abds}\} $\cup$ Store\\ 
\hline

4& {\tt \Rs} $\Diamond$  Answ $\Diamond$ Store& inconst(Store)&
failure\\
\hline

5a&{\tt \{p(\vect{X}) $\leftarrow$ Abds\}} $\cup$ {\tt \Rs}&  &{\tt \Rs}
$\Diamond$  Answ $\Diamond$ \\
& $\Diamond$  Answ $\Diamond$ Store & &\{subs({\tt
  p(\vect{X}) $\leftarrow$ Abds},Answ$_p$)\} \\ 
& & & $\cup$  Store\\
\hline

5b&{\tt \{p(\vect{X}) $\leftarrow$ Abds\}} $\cup$ {\tt  \Rs} &  &{\tt \Rs}
$\Diamond$ {\tt \{p(\vect{X}) $\leftarrow$ Abds\}} $\cup$  Answ $\Diamond$ \\
&$\Diamond$  Answ $\Diamond$ Store & &\{not\_subs({\tt
  p(\vect{X}) $\leftarrow$ Abds},Answ$_p$)\} \\ 
& & & $\cup$  Store\\
\hline
\end{tabular}
    
    \caption{Inference rules of constraint procedure.}
    \label{fig:constr}
  \end{center}
\end{figure}

Rules 1a, 1b and 2 are as before. Rule 3 adds a new constraint to the
constraint store. Rule 4 stops the derivation with failure when the
store is inconsistent. Rule 5 processes a new answer lemma. It is a
non deterministic rule. 5a handles the alternative where it is
enforced that the new answer lemma is subsumed by the existing answers
(Answ$_p$). The lemma is deleted and the subsumption constraint is
added to the store. 5b handles the case where it is enforced that the
answer is not subsumed by the previous one.  It is added to the
answers and the negation of the subsumption constraint is added to the
store.

\paragraph{Correctness and termination.}

This proof procedure preserves the properties about correctness
and completeness of the abductive one. In terms of the inferences it
makes, the difference is that it uses constrained facts {\tt
  \{p(\vect{X}) $\leftarrow$ Abds\}} instead of facts {\tt
  p(\vect{t})} and that it delegates the processing of the calls to
the abducibles to the constraint solver. Hence the inferences it makes
are correct and completeness is preserved. Termination remains ensured
if the number of answers for each program predicate remains finite and
consistency checking is terminating. A {\tt not(subsumed(\ldots))}
constraint stating that the new answer is not subsumed by the previous
ones is added to the constraint store each time that a new answer is
added to the answer set. With $n$ the size of the domain, the number
of distinct atoms in the model of a $m$-ary predicate is limited to
$m^n$. Hence, if more than $m^n$ answers are added for the same
predicate, an inconsistent store will be reached. The consistency
check has to verify that a pre-interpretation exists which satisfies
all constraints. As the number of different pre-interpretations is
finite and the number of constraints is finite, its termination can be
ensured.

\paragraph{Control and consistency checking.}

The best way to handle the choice offered by rule 5 is through
enumeration and backtracking where preference is given to rule 5a as
it leads to the smallest answer set, hence to the shortest derivation.

In the abductive algorithm the choice implies a commitment for a
particular component of the pre-interpretation. In this algorithm the
choice does not imply such a direct commitment. However, adding a non
redundant constraint reduces the number of pre-interpretations that
satisfy all constraints; it is a kind of indirect commitment.

The solver which has to verify the consistency plays a crucial role.
We have explored two alternatives. The first approach (abductive
solver) abduces the components of the pre-interpretation as needed
during the verification of the constraints. Backtracking is triggered
when a constraint is violated. (This is similar to the strategy in the
abductive algorithm, but at the level of the constraint checking.)
Note that the constraint checking can be incremental; each time a new
constraint is added, the search starts from the partial
pre-interpretation satisfying all previous constraints.

The second approach (finite domain solver) encodes the search for a
pre-interpretation as a finite domain problem. A finite domain
variable ranging over the domain of the pre-interpretation is
associated with the terms occurring in the constraints and boolean
variables are used to express the equality between the
pre-interpretation of different terms. We sketch the encoding using
the even-odd example.  Let ${\cal D}$ be the domain of the
pre-interpretation. A finite domain variable $D_t$ ranging over ${\cal
  D}$ represents the pre-interpretation of a term $t$ and boolean
variables $B_{t_1=t_2}$ indicates whether or not the terms $t_1$ and
$t_2$ have the same pre-interpretation. Such boolean variables are
linked to the domain variables through definitions $B_{t_1=t_2}
\leftrightarrow D_{t_1} = D_{t_2}$ which ensure propagation of new
information.  Consider the constraint {\tt false $\leftarrow$
  abduce(0,X), abduce(s(0),X)}. To handle it we introduce finite
domain variables $D_0$ and $D_{s(0)}$. We can translate the constraint
to {\tt false $\leftarrow$ $D_0$ = X, $D_{s(0)}$ = X} or, after
elimination of {\tt X}: {\tt false $\leftarrow$ $B_{0=s(0)}$} or
$B_{0=s(0)} = 0$. To express the connection between $0$ and $s(0)$, we
add for all $d \in {\cal D}$ the constraint $B_{0=d} \le
B_{s(0)=s(d)}$\footnote{Or equivalently $B_{0=d} \rightarrow
  B_{s(0)=s(d)}$.}. Note that this implies the creation of finite
domain variables $D_{s(d)}$.  Now consider the constraint {\tt
  subsumed(even(Y) $\leftarrow$ abduce(s(s(0)),Y),\{even(X)
  $\leftarrow$ abduce(0,X),\})}.  It contains a new term $s(s(0))$, so
a domain variable $D_{s(s(0))}$ is created and it is linked with
$D_{s(0)}$ by adding for all $d \in {\cal D}$ the constraint
$B_{s(0)=d} \le B_{s(s(0))=s(d)}$.  The subsumption constraint is
expressed as $B_{s(s(0))=0}=1$. (Its negation is represented as
$B_{s(s(0))=0}=0$.)  This translation ensures that all choices which
are made are immediately propagated.

\subsection{Failure analysis and symmetries}
\label{sec:failanalyse}

Consider the abductive system and the constraint system with the
abductive solver. Chronological backtracking is triggered by the
derivation of a {\tt false$\leftarrow$} clause. However, not all the
components of the pre-interpretation abduced so far necessarily have
contributed to its derivation.  So, chronological backtracking may
result in thrashing. The amount of backtracking can be substantially
reduced. In the context of the abductive system, a simple approach is
to associate with each clause the set of abductive facts used in its
derivation. In each derivation step, the set associated with the new
clause is the union of the sets of the two parent clauses. When
abducing a new fact, the associated set is the abduced fact itself. In
this way, when {\tt false$\leftarrow$} is derived, one obtains an
associated {\em conflict set} identifying the abduced facts used in
the derivation of the clause. Backtracking is then directed to the
last abduced fact in the set. To support also the derivation of
secondary conflict sets, the technique of intelligent backtracking
\cite{IB} is used. With $S_1 \wedge \{abduce_f(\vect{s},d1)\}$ a
conflict set which backtracks to the generator of
$abduce_f(\vect{s},\_)$, the conflict set $S_1 \wedge
\{abduce_f(\vect{s},d1)\}$ is stored with the generator of this
abductive component. If it happens that all possible assignments for
that component get rejected, then one obtains a set of conflicts which
can be formalised as: $S_1 \wedge \{abduce_f(\vect{s},d1)\}
\rightarrow false, \ldots ,S_n \wedge \{abduce_f(\vect{s},dn)\}
\rightarrow false$. Applying hyper-resolution on these clauses and
$abduce_f(\vect{s},d1) \vee \ldots \vee abduce_f(\vect{s},dn)$, which
expresses that there must be a domain element assigned to the term
$f(\vect{s})$, one obtains the secondary conflict $S_1 \wedge \ldots
\wedge S_n \rightarrow false$ and one can backtrack to the most recent
abducible in that set. Note that a trade-off between the time lost in
rediscovering the same conflict set and the time and space lost in
storing and checking previous conflict sets has to be made. It is done
by storing the conflict set with the generator of the most recent
abduced fact present in the conflict set.  This approach avoids
inefficiencies in accessing relevant conflict sets. However, if, due
to another conflict, one backtracks beyond that generator, then the
information about the conflicts stored with it is lost.
Notwithstanding that this conflict set still can have potential use for
future pruning.

The conflict sets obtained in this way are not optimal. In fact, as Peltier
\cite{Peltier98} points out, it is not feasible to compute optimal
conflict sets: if there is no model for the given domain size, the
optimal conflict set is empty. However further improvements are
feasible. For example, consider the clause {\tt false $\leftarrow$ t1 =
  t2, Y = t3}. It leads to the clause 
{\tt false$\leftarrow$} whenever the pre-interpretations of t1 and t2
are equal. However, our approach will also include the components used in
computing the pre-interpretation of t3. Improved failure analysis
could be achieved by applying substitutions. Indeed for any expression
E, the (pre-) interpretation of $E, X = t$ for which $X$ does not
occur in $t$ is equal to the (pre-) interpretation of $E\{X/t\}$. 
The abductive constraint solver uses this rule to simplify {\tt false
  $\leftarrow$ Abds} constraints. The abductive procedure does not use
it as it performs abstract compilation as a pre-processing step and the
integration of this optimisation would require a complete redesign of
the system (one should keep track of the used components of the
pre-interpretation at the level of individual terms instead of at the
clause level). The abductive constraint solver also sharpens the
conflict set of violated subsumption constraints: it selects an
argument for which the subsumption constraint is violated and returns
as conflict set the components used in evaluating that argument.

As noted by Peltier \cite{Peltier98}, applying a permutation of the
domain on a partial (pre-) interpretation yields an isomorphic (pre-)
interpretation. It can be extended into a model iff the original one
can be extended in a model. In particular, if $S$ is a conflict set, so
is $S\{d_1/d'_1, \ldots, d_n/d'_n\}$ with $\{d'_1,\ldots,d'_n\}$ a
permutation of $\{d_1,\ldots,d_n\}$. There is again a trade-off
between the time lost in rediscovering an isomorphic conflict and in
storing and using such conflicts to prune the search. We follow what
we understand to be the approach in \cite{Peltier98}: when a conflict
set is found and the system backtracks and considers the next
candidate, it is checked whether that candidate has a subset which is
isomorphic to the conflict set which triggered the backtracking. For
example, assume a conflict $\{abduce_f(d1,d1),abduce_a(d2)\}$ is
derived and the enumeration modifies the second component in
$abduce_a(d3)$. The new partial pre-interpretation contains the set
$\{abduce_f(d1,d1),abduce_a(d3)\}$ which is isomorphic to the original
one under the permutation $\{d3/d2,d2/d3\}$ and is rejected.
Note that $\{ abduce_f(d2,d2), abduce_a(d1)\}$ is also isomorphic to
the original conflict set. While it can be part of a candidate still
to be explored in the search space, this conflict set is not stored
for future pruning. 

\section{Alternative approaches}
\label{sec:alt}

\paragraph{Model generation.}
The logic program and the clause {\tt false $\leftarrow$ query} can be
considered as a logical theory. A model of this theory is a proof that
the query fails. There exist general purpose tools for generating
models of logical theories. FINDER \cite{FINDER1,FINDER}, written in
C, is such a tool; it takes as input a set of clauses in a many-sorted
first order language, together with specifications of finite
cardinalities of the domains for the sorts, and generates
interpretations on the given domains which satisfy all the clauses
\cite{FINDER1}. A basic difference with our approach is that it not
only enumerates the pre-interpretation but also the interpretation
(the mapping from the atoms to true or false). Also, it is not goal
directed; the system checks whether all ground instances of all
clauses are true in the candidate interpretation. If a clause instance
is found which is false, then the components of the interpretation
which have been used in the evaluation make up what we called the
conflict set and are used to direct the search (using hyper-resolution
as described in Section \ref{sec:failanalyse} to derive secondary
conflicts). Another difference is that FINDER stores conflict sets
permanently (unless they are too big) and uses elaborate algorithms to
exploit them efficiently in pruning the search. FINDER reports a model
when it discovers an interpretation which is true for all clause
instances.  FINDER uses a typed language and can only handle binary
predicates and functors. To overcome that restriction we used a
special encoding for predicates and functors of higher arity. For
example, with $n$ the size of domain $D$ and $f/3$ a ternary functor,
we introduce a domain $D_h$ with cardinality $n \times n$, a binary
functor $f_h:D \times D \rightarrow D_h$, and a binary functor
$f_b:D_h \times D \rightarrow D$. Then we replace every occurrence
$f(t_1,t_2,t_3)$ by $f_b(f_h(t_1,t_2),t_3)$. In addition $f_h/2$
should have a different value for each different input and FINDER
should not backtrack over the choices made for $f_h/2$. This can be
achieved with declaring $f_h$ injective and ensuring that the choice
points for $f_h/2$ are created first.  \cite{FINDER} states that the
order in which functors are declared is important and that the first
declared ones change least rapidly during the backtracking.  The first
results we obtained with FINDER \cite{PLILP98} were rather poor. The
results reported in the current paper are much better. They are
obtained with the ordering which places the special functor $f_h$
first, then the constants, then the other functors and finally the
predicates.

In a recent paper, Peltier \cite{Peltier98} presents a new system
FMC$_{\rm ATINF}$ which claims to do a better failure analysis than FINDER and
SEM \cite{SEM}, and exploits symmetries to further improve the
pruning of the search space. As with other finite model builders for first
order logic, it enumerates the full interpretation, not only the
pre-interpretation. Our understanding is that the concept of {\em
  covering refutation} used to prune the search is very similar to our
use of intelligent backtracking (which we added to our system
described in \cite{PLILP98} after we learned about the work of Peltier).

\paragraph{Regular approximations.}
Within the context of program analysis, the most obvious approach to
prove failure is to add a clause {\tt shouldfail(\vect{X})
  $\leftarrow$ query(\vect{X})} and to use one or another kind of type
inference to show that the success-set of {\tt shouldfail(\vect{X})}
is empty. A typical representative of such systems is described in
\cite{Gal94}; it computes a regular approximation of the program.
Roughly speaking, for each argument of each predicate, the values it
can take in the success-set\footnote{The set of ground atoms which are
  logical consequences of the program.}  are approximated by a type (a
canonical unary logic program). Failure of the query is proven if the
types of {\tt shouldfail(\vect{X})} are empty. Also set based analysis
\cite{Heintze-sc-anal-92} can be used to approximate the success-set.
Set-based analysis originates from \cite{Mishra-84}; it was then
studied (improved and implemented) in \cite{Heintze-sc-anal-92}.  The
tool that we use is a composition of inference of a directional type
(as in \cite{CharatonikPodelski-SAS98}, based on set-based analysis)
with the theorem prover SPASS \cite{Weidenbach97jar}. 

\paragraph{Program specialisation.}
One could also employ program transformation, and more specifically
program specialisation techniques to prove failure of the query. If
for the given query, the program can be specialised in the empty
program, then the query trivially fails. A technique which has almost
the same power as transformations based on the fold/unfold approach is
conjunctive partial deduction \cite{LeuschelDeSchreyeDeWaal:JICSLP96,CPD}.
By specialising conjunctions of atoms instead of single atoms, it can
achieve substantially better results than other specialisers. For
example it can specialise the even/odd program into the empty program
for our example query.

\section{Experiments}
\label{sec:exp}

Table \ref{tab:programs} gives details about the benchmarks\footnote
{The code is available at http://www.cs.kuleuven.ac.be/\~{}henkv/pre}.
Besides the name, the table gives the number of clauses (without the
query), the number of predicates, the size of the domain, the size of
a pre-interpretation, the number of different pre-interpretations
(including symmetric ones), the size of an interpretation (the number
of atoms to be mapped to true or false), and the number of different
interpretations (for a fixed pre-interpretation).

\begin{table}[htbp]
  \begin{center}
    \leavevmode
    \begin{tabular}{|l|r|r|r|r|r|r|r|}
      \hline \hline
      \multicolumn{1}{|c|}{name}&\multicolumn{1}{c|}{\#clauses}&
      \multicolumn{1}{c|}{\#pred}&\multicolumn{1}{c|}{size(dom)}&
      \multicolumn{1}{c|}{size(pre)}&
      \multicolumn{1}{c|}{\#pre}& 
      \multicolumn{1}{c|} {size(int)}& \multicolumn{1}{c|} {\#int/pre}\\
      \hline
      odd\_even    & 3 & 2 & 2 & 3 & $2^{3}$ & 4& $2^{4}$\\
      wicked\_oe   & 4 & 3 & 2 & 10 & $2^{10}$ & 10& $2^{10}$\\
      \hline
      appendlast   & 4 & 2 & 3 & 12 & $3^{12}$& 13& $2^{13}$\\
      reverselast  & 4 & 2 & 3 & 12 & $3^{12}$& 13& $2^{13}$\\
      nreverselast & 6 & 3 & 5 & 28 & $5^{28}$& 150& $2^{150}$\\
      schedule     & 12 & 6 & 3 & 12 &$3^{12}$& 24& $2^{24}$\\ 
      \hline
      multiseto    & 7 & 1 & 2  & 7 & $2^{7}$& 4& $2^{4}$\\
      multisetl    & 4 & 2 & 2  & 7 & $2^{7}$& 12& $2^{12}$\\
      \hline
      blockpair2o  & 15 & 3 & 2 & 19 & $2^{19}$& 12& $2^{12}$\\
      blockpair3o  & 15 & 3 & 2 & 36 & $2^{36}$& 20& $2^{20}$\\
      blockpair2l  & 14 & 5 & 2 & 19 & $2^{19}$& 32& $2^{32}$\\
      blockpair3l  & 14 & 5 & 2 & 36 & $2^{36}$& 40& $2^{40}$\\
      blocksol     & 14 & 5 & 2 & 19 & $2^{19}$& 32& $2^{32}$\\
      \hline
      BOO019-1        & 4 &  1 & 3 & 32 & $3^{32}$& 9 &  1 \\
      \hline \hline
    \end{tabular}
    \caption{Properties of benchmark programs}
    \label{tab:programs}
  \end{center}
\end{table}

{\tt odd\_even} is a trivial example about even and odd numbers. {\tt
  wicked\_oe} is an extension which adds a call to each clause and 4
functors which are irrelevant for success or failure.  It allows us to
see whether failure analysis is accurate enough to achieve the same
level of pruning as in {\tt odd\_even}. {\tt appendlast, reverselast,
  nreverselast} and {\tt schedule} are small but hard examples.  On
one hand, they illustrate the use of integrity constraints to express
program properties; on the other hand, they have circulated as
challenging problems for program specialisers which should be able to
specialise them into the empty program. The query for {\tt appendlast}
expresses the integrity constraint that appending a list ending in
{\tt a} cannot result in a list ending in a {\tt b}. The query for
{\tt reverselast} expresses that reversing a list with the
accumulating parameter initialised as {\tt [a]} cannot result in a
list ending in a {\tt b}. The query for {\tt nreverselast} expresses
naive reverse applied on a list beginning with an {\tt a} cannot
result in a list ending in a {\tt b}. Finally, the {\tt schedule}
program is a program that attempts to transpose successive positions
in a list of elements until a configuration is reached with two
successive {\tt c} elements. The query expresses the integrity
constraint that such configuration cannot be reached from a
configuration consisting of one {\tt c} followed by one or more {\tt
  n}.  {\tt multiseto} and {\tt multisetl} are programs to check the
equivalence of two multisets. While the first uses a binary operator
``o'' to build sets, the second uses a list representation and
auxiliary predicates to manipulate the list.  The others are typical
examples from a large set of planning problems reasoning on multisets
of resources.  The first two use the ``o'' representation for the
multiset, the next two the ``l'' (list) representation. {\tt blockpair2o} and
{\tt blockpair2l} omit the for success or failure irrelevant argument
for collecting the plan (and have 6 functors less). {\tt blocksol} is
there to show what happens when the query does not fail. It uses the
list representation and also omits the argument collecting the plan.
Finally, {\tt BOO019-1} is an axiomatisation of a ternary boolean
algebra, a typical problem from the theorem proving community taken
from the TPTP library \cite{TPTP}). Its only predicate is equality,
whose interpretation can be fixed to the identity (so the number of
different interpretations is only 1 instead of $2^9$).

The abductive system AB uses the control as described in Section
\ref{sec:abductive} and is augmented with the intelligent backtracking
and symmetry checking as described in Section \ref{sec:failanalyse}.
The constraint system gives lowest priority to rule 5 which create a
choice point. Rule 4 which checks for consistency of the constraints
is activated each time a new constraint is generated. The system with
the abductive solver (ABCS) also uses intelligent backtracking and is
able to derive more accurate conflict sets than AB. The system with
the finite domain solver (FDCS) does not apply intelligent
backtracking as this is difficult to integrate with the standard
pruning techniques of finite domain solvers \cite{hent89}. Both
constraint systems eliminate only the most obvious symmetry in the
search space (when starting the enumeration with {\tt abduce$_a$(X)}
they will consider only one domain value for {\tt X}).

The abductive systems are implemented in Prolog. The queries have been
executed with MasterProlog on a SUN sparc Ultra-2. The constraint
system FDCS is also written in Prolog, uses the SICSTUS finite domain
solver and was running under SICSTUS Prolog \cite{SICSTUS} on the same
machine. ABCS was also running under SICSTUS Prolog.  FINDER is
implemented in C and was also running on a SUN sparc Ultra-2.  Regular
approximations (RA) were computed with a system due to John Gallagher,
conjunctive partial deduction (CPD) with a system due to Michael
Leuschel. Witold Charatonik was so kind to run our examples on his
tool for set based analysis (SBA) described in Section \ref{sec:alt}.
Finally, Nicolas Peltier was so kind to run our examples on the
FMC$_{\rm ATINF}$ system \cite{Peltier98} under a SUN4 ELC.

Table \ref{tab:times} gives the times for the various systems. The
times are in seconds, unless followed by H, in which case it is in
hours. The notation $>x$H means no solution was found after $x$ hours;
M means out of memory.  For CPD, RA and SBA, we do not give times as
these systems do not perform an exhaustive search but a standard
analysis of the given program. {\em yes} means failure was proven. As
{\tt blocksol} does not fail, it was not run on these systems.

Table \ref{tab:back} gives the number of backtracks.
For the constraint systems, two numbers are given, \#bckt is the
number of backtracks inside the solver, i.e.\ those with respect to the
pre-interpretation, \#Tbckt is the number with respect to the choices made
regarding tabling.

\begin{table}[h]
  \begin{center}
    \leavevmode
    \begin{tabular}{|l|r|r|r|r|r|r|r|r|}
      \hline 
      \multicolumn{1}{|c|}{name}& 
      \multicolumn{1}{|c|}{AB} &
      \multicolumn{1}{|c|}{FDCS}&
      \multicolumn{1}{|c|}{ABCS}&
      \multicolumn{1}{|c|}{FINDER}&
      \multicolumn{1}{|c|}{FMC}&
      \multicolumn{1}{|c|} {CPD}&
      \multicolumn{1}{c|} {RA} &
      \multicolumn{1}{c|} {SBA}\\
      \hline 

      odd\_even &  0.01 &  0.00 & 0.00  &0.02    & 0.00  &yes &yes&yes\\
      wicked\_oe&  0.07 &  0.00 & 0.00  &0.02    & 0.01  &yes &yes&yes\\
      \hline 
     appendlast &  0.53 & 0.45  & 0.09  & 0.17   & 45.21 &yes &no &yes\\
    reverselast &  0.38 & 3.70  & 0.94  & 0.17 & 10.79 &no  &no &yes\\
   nreverselast & 5.87H &$>$19H &$>$19H & $>$16.5H & $>$900  &yes &no &no\\
      schedule  &  0.10 &  0.31 & 0.07  & 0.03   & 0.15  &no  &no &yes\\
     \hline 
     multiseto  &  0.04 &  0.04 & 0.02  &0.02    & 0.02  &no  &yes&yes\\
     multisetl  &  0.01 &  0.06 & 0.03  &0.02   & 0.08  &yes & no&yes\\ 
     \hline 
    blockpair2o &  1.83 &  0.38 & 0.11  & 0.08  & 7.31  &no  &no &no\\
    blockpair3o & 7.60  & 0.42  & 0.14  & 0.18  & $>$900  &no  &no &no\\
     blockpair2l&  2.83 & 2.36  & 1.17  & 0.05  & 204.9 &no  &no &no\\
     blockpair3l& 29.24 & 2.49  & 1.34  & 0.12  & M  &no &no  &no\\
      blocksol  &200.78 & 7.7H &2558.58 &1896.3 & $>$900  &-   &-  &-\\
      \hline
      BOO019-1  & 1.20  & 4.34  & 0.14  & 0.03 & 0.06  & no & no& no \\
      \hline 
    \end{tabular}
    \caption{Times.}
    \label{tab:times}
  \end{center}
\end{table}

\begin{table}[h]
  \begin{center}
    \leavevmode
    \begin{tabular}{|l|r|r|r|r|r|r|r|r|r|r|}
      \hline 
      \multicolumn{1}{|c|}{name}& 
      \multicolumn{1}{|c|}{AB} &
      \multicolumn{2}{|c|}{FDCS}&
      \multicolumn{2}{|c|}{ABCS}&
      \multicolumn{1}{|c|}{FINDER}&
      \multicolumn{1}{|c|}{FMC}\\
      \hline 
      \hline 
      &\multicolumn{1}{|c|}{\#bcktr} &
      \multicolumn{1}{|c|}{\#bcktr}&
      \multicolumn{1}{|c|}{\#Tbcktr}&
      \multicolumn{1}{|c|}{\#bcktr}&
      \multicolumn{1}{|c|}{\#Tbcktr}&
      \multicolumn{1}{|c|}{\#bcktr} &
      \multicolumn{1}{|c|}{\#bcktr} \\
      \hline 
      odd\_even &      4 &  0 & 0  &0    & 0   & 1       &3\\
      wicked\_oe&     64 & 0  & 0  & 0   & 0   & 8       &52\\
      \hline 
     appendlast &  43    & 24 & 1  & 55  & 1   & 618     &110019\\
    reverselast & 30     & 68 & 2  & 303 & 2   & 614     &23445\\
   nreverselast & 190170 & ?  & ?  &  ?  & ?   &$>5.10^7$& ?\\
      schedule  &  24    & 13 & 1  &106  & 1   & 37      &497\\
     \hline 
     multiseto  &  10    &   7& 0  &30   & 0   & 0       &104\\
     multisetl  &  3     & 6  & 1  & 21  & 1   & 12      & 469\\ 
     \hline 
    blockpair2o &  17    & 25 & 0  &49   & 0   & 262     & 5567\\
    blockpair3o & 56     & 25 & 0  &51 & 0     & 879     &?\\
     blockpair2l&  28    &3943& 2  &2733 & 2   & 68      &91404\\
     blockpair3l& 130    &4009& 2  &2737 & 2   & 366     &?\\
      blocksol  & 3615&1396146& 385&1970544&169& 4007523 &?\\
      \hline
    BOO019-1    & 72     & 4  & 0  & 34  & 0   & 14      & 33 \\
      \hline 
    \end{tabular}
    \caption{Amount of backtracking.}
    \label{tab:back}
  \end{center}
\end{table}

\subsection{Discussion}

One should refrain from comparing results for individual examples. A
different order over the choice points can give a very different
result. This is definitely so for FINDER, which imposes an almost
static ordering over the choice points. The order used in
\cite{PLILP98} was giving much worse results. The abductive system AB
determines the order dynamically. Still, a small change in the
heuristics for selecting the next rule application can result in a
different order over the choice points and in substantially different
results. For example, a variant of the AB system solved {\tt
  blockpair3l} in 2.60s with 18 backtracks (but was doing worse when
considering the whole benchmark suite).

The effects of pruning based on symmetries in the AB system is not
reported in the tables. Of the problems with 2 element domains, it
triggers only one pruning step in {\tt blocksol} which has to search
the whole solution space. It causes one pruning step in each of the
problems with 3 element domains, but makes the difference on {\tt
  nreverselast} where a 5 element domain is needed to construct a
model. There, it causes 120 pruning steps in the search space of the
AB system.

Comparing the twin problems {\tt odd\_even} and {\tt wicked\_oe}, {\tt
  blockpair2o} and {\tt blockpair3o}, {\tt blockpair2l} and {\tt
  blockpair3l}, we observe that FDCS with the finite domain solver is
almost not distracted by the (for the failure) irrelevant extra
functors. Apparently, its control strategy is such that those extra
functors are enumerated as the last ones, when the more constrained
functors already received a correct assignment in the domain of the
pre-interpretation. Also the failure analysis of the abductive solver
ABCS turns out to be very accurate and the amount of backtracking is
almost unaffected by the extra functors.  The failure analysis of the
abductive system AB includes those functors in conflict sets so that
the backtracking becomes less accurate and more backtracks occur
before a solution is found. This also holds for the model generators
FINDER and FMC$_{\rm ATINF}$.

The abductive system AB is the only system solving all problems and is
doing very well in terms of number of backtracks (apart from {\tt
  wicked\_oe} and {\tt BOO019-1}). Its implementation is very
straightforward (linear lists for clauses, abduced facts and tabled
answers), so there is a lot of room for improving its speed.  As a
consequence it is often slower than the constraint systems. The
latter, also implemented in Prolog, use more elaborate data
structures.

Both constraint systems FDCS and ABCS are doing pretty well. Although
they are also implemented in Prolog, they use more elaborate data
structures and are often faster than AB, even when they need more
backtracks. Their performance degrades when they have to backtrack
frequently over their decision with respect to the subsumption of new
answers (\#Tbcktr). This is a major weakness. The more they have to
backtrack over the subsumption decisions, the more the size of their
search space becomes close to that of the model generators FINDER and
FMC$_{\rm ATINF}$. A problem is that they have no control over the
order of these choice points and cannot do better than chronological
backtracking over all possibilities. This problem is prominent in {\tt
  blocksol} where they have to search the whole space of
interpretations of size $2^{19} \times 2^{32}$ whereas AB searches in
the space of pre-interpretations of size $2^{19}$. The poor
performance in this problem is not a real drawback. The use of the
model generator should be combined with the use of a theorem prover
which should be able to find a plan for {\tt blocksol}. We suspect
this problem is at the basis of their failure on the {\tt nreverselast}
problem where the domain size is 5. A positive point is that answer
lemmas can contain variables, in which case they cover several ground
answers. This limits somewhat the number of different answer sets
which have to be considered during the search. (FINDER needs much more
backtracks on the {\tt blocksol} problem.)

Whereas our original results \cite{PLILP98} confirmed those of
\cite{Peltier98} that FMC$_{\rm ATINF}$ most of the time outperforms
FINDER, more fine tuning of the FINDER input reversed the picture. Its
implementation in C and its use of specialised data structures pays of
on the class of problems we consider. It is fast on all but the hard
problems. In {\tt blocksol} it is beaten by AB (over 4 million
backtracks against 3615 for AB) and it was stopped on {\tt
  nreverselast} after 50 million backtracks. FINDER is pre-processing
short clauses. Likely this eliminates a lot of candidate solutions (it
solves {\tt multiseto} without backtracking). FMC$_{\rm ATINF}$, which
has no such pre-processing, is unable to solve several problems, in
particular the planning problems with a list representation. The
latter problems are those where the the reduction of search space by
our approach (the space of pre-interpretations versus the space of
interpretations  ---see last column in Table \ref{tab:programs}---) is
largest. 

The advantage for our systems that they search the smaller space of
pre-interpretations disappears on the TPTP problem {\tt BOO019-1} and
similar problems. Hence the first order model generators do as well or
better in terms of number of backtracks and, due to their fine tuned
C implementations, outperform our systems in speed.

Conjunctive partial deduction can handle some of the problems which
are difficult for us, but cannot handle any of the planning problems.
Computing regular approximations is fast, but it can show failure of
the most simple problems only. The set based analyser is more precise
and fails only on the planning problems,  {\tt nreverselast} and {\tt
  BOO019-1}.

\section{Conclusion}
\label{sec:conc}

For definite logic programs, we have addressed the problem of proving
that certain queries cannot succeed with an answer. A problem which is
particularly relevant when the query does not fail finitely. We have
developed two new approaches which aim at searching a model of the
program in which the query is false.  We have performed some
experiments using (rather small) example programs and queries which do
not terminate\footnote{These programs also loop when using tabulation
  or when executing bottom-up after a magic set transformation.}.  We
also did a comparison with other approaches which could be used to
tackle this problem: general purpose model generation tools, the use
of type inference, and the use of program specialisation. In the case
of type inference, the approach is in fact also to compute a model.
However, the chosen model is the one which best reflects the type
structure of the program. If the query happens to be false in this
model, then failure is shown. Also in the case of program
specialisation, showing failure is a byproduct of the approach: for
some queries, the program happens to be specialised into the empty
program.

Abduction is a very powerful and general problem solving technique. It
was pretty easy to formulate the problem of searching a
pre-interpretation such that the query is false in the least model
based on it as an abductive problem and to use a general purpose
abductive procedure\cite{SLDNFA}. But we quickly realised that we had
almost no control over the search for a solution. Our first approach
was to built a special purpose abductive procedure for definite
programs which employs tabulation and which hard wired the constraints
that pre-interpretation of functors are total functions. The idea
behind the proof procedure is to use a top-down evaluation strategy
---abducing a part of the pre-interpretation only when needed in
evaluating the query--- and to prevent looping by the use of
tabulation. Experiments confirmed our intuition that it was important
to delay the abduction of new components in the pre-interpretation as
long as possible (to propagate all consequences of what was already
abduced to check whether it was part of a feasible solution).  After
adding failure analysis to improve upon chronological backtracking as
in systems as FMC$_{\rm ATINF}$ \cite{Peltier98} and FINDER
\cite{FINDER1,FINDER}, the system is doing quite well. It outperforms
FMC$_{\rm ATINF}$ in speed and number of backtracks. Compared with
FINDER, it typically needs much less backtracks, though it can only
beat FINDER in speed for a couple of hard problems. 

We also explored a variant which treats the definition of the
pre-interpretation as constraints.  This allows to delay the decisions
up to the point were answers had to be tabled: at such a point one
needs to know whether the answer is new or not. Still we do not fix the
pre-interpretation at such a point but formulate constraints on the
pre-interpretation, using a solver to check the existence of a
pre-interpretation satisfying all constraints.  We experimented with a
finite domain solver and with an abductive solver.  We obtained good
results; however the systems start to slow down when a lot of
backtracking over the decisions with respect to new answers being
subsumed by the existing ones is needed. The number of possible
backtracks quickly goes up with the arity of the predicates, as the
example {\tt blocksol}, where the query does not fail, illustrates. It
also increases quickly with the size of the domain needed to show
failure e.g.\ {\tt nreverselast}. Unless one finds some heuristics to
control the order of choice points, or some knowledge to do better than
the chronological backtracking over these choice points, the abductive
system seems more promising.

Our experiments indicate that our approach is a better basis for
proving failure of queries over definite programs than applying
general purpose model generators such as FINDER and FMC$_{\rm ATINF}$.
Searching the space of pre-interpretations for a pre-interpretation
such that the query is false in the least model based on it requires
on average much less backtracking than searching the larger space of
interpretations. Searching in the smaller space of pre-interpretations
has a cost: the query needs to be evaluated in the least model.
Tabulation turned out to be a very effective approach which keeps the
cost of the query evaluation at acceptable levels. (As the program
abstracted under the pre-interpretation is a DATALOG program, also
loop checking can ensure termination of the query evaluation. At some
point we experimented with this and got a very large slow down as the
number of derived clauses substantially increased.)

Our approach is also more powerful than type inference
based on Regular Approximations. Conjunctive Partial Deduction and Set
Based Analysis turn out to be quite powerful on some classes of
problems but cannot solve any of the planning problems.

A limitation of our approach, but also of the model generators is that
they cannot prove failure if the query is only false in a model based
on an infinite domain. For example {\tt less(N,s(N)) $\leftarrow$} and {\tt
  less(N,s(M))  $\leftarrow$ less(N,M)} and the query {\tt $\leftarrow$
  less(N,M),less(M,N)}. Also set based analysis and conjunctive partial
deduction are unable to prove failure of this query.

In a broader context, this paper makes contributions to the following
topics:
\begin{itemize}
\item A (first)  study of methods to prove (infinite) failure of
  definite logic programs. 
\item The development of a proof procedure which combines tabulation
  with abduction and of a constraint based procedure which treats the
  abducibles as constraints and uses a constraint solver to check the
  existence of a solution for the abducibles. Also the latter
  procedure uses tabulation.
\item A better understanding of the power and limitations of
  abduction. While very expressive, our findings suggest that
  abductive procedures need to be augmented with ``background''
  knowledge to direct the search for abductive solutions. Simply
  specifying the properties of an abductive solution as an integrity
  constraint cannot provide sufficient guidance to the search for a
  solution. It is interesting to observe that background knowledge is
  also often the key to success in Inductive Logic Programming which
  makes use of inductive procedures which are in more than one aspect
  ``twins'' of abductive procedures \cite{Ade95}.
\item The further development of model based program analysis.
  \cite{Gal} showed that model based program analysis implicitly
  introduced in \cite{CD95} is also an excellent method for type
  inference. In \cite{Gal} it was shown that there exist
  pre-interpretations which encode various other declarative
  properties of programs. Our work takes this work one step further by
  developing methods for automatically constructing a
  pre-interpretation which expresses a particular program property (or
  integrity constraint) expressed as a query which should fail.
\end{itemize}

\section*{Acknowledgements}

The author are grateful to the reviewers for the detailed comments and
the many suggestions for improvement.
M. Bruynooghe is supported by FWO-Flanders. This research was
supported by the EC (a grant from the HCM project ERBCHBGCT9303365 to
D.A. de Waal), by the INTAS project 93-1702 {\em Efficient Symbolic
  Computing},  by the GOA projects {\em Non-Standard Application of Abstract
  Interpretation} and {\em LP+: a Second Generation Logic Programming
  Language} and by the FWO projects {\em Linguaduct} and {\em
  Knowledge Representation and Computation in Open Logic Programming}.

\begin{small}


\begin{thebibliography}{10}

\bibitem{Ade95}
H.~Ade and M.~Denecker.
\newblock Abductive inductive logic programming.
\newblock In C.S. Mellish, editor, {\em Proc. of the International Joint
  Conference on Artificial Intelligence}, pages 201--1209. Morgan Kaufman,
  1995.

\bibitem{apt:iclp89}
K.R. Apt, R.N. Bol, and J.W. Klop.
\newblock On the safe termination of {Prolog} programs.
\newblock In {\em Proc. ICLP'89}, pages 353--368, Lisbon, June 1989. MIT Press.

\bibitem{bol:tcs}
R.N. Bol, K.R. Apt, and J.W. Klop.
\newblock An analysis of loop checking mechanisms for logic programs.
\newblock {\em Theoretical Computer Science}, 86(1):35--79, August 1991.

\bibitem{JLP92}
M.P. Bonacina and J.~Hsiang.
\newblock On rewrite programs: semantics and relationship with {P}rolog.
\newblock {\em Journal of Logic Programming}, 14(1\&2):155--180, October 1992.

\bibitem{s-semantics}
A.~Bossi, M.~Gabbrielli, G.~Levi, and M.~Martelli.
\newblock The s-semantics approach: Theory and applications.
\newblock {\em J. Logic Programming}, 19/20:149--197, 1994.

\bibitem{Dima2}
D.~Boulanger and M.~Bruynooghe.
\newblock A systematic construction of abstract domains.
\newblock In {\em Proceedings SAS'94 (Static Analysis Symposium)}, LNCS 864,
  pages 61--77. Springer-Verlag, 1994.

\bibitem{Dima1}
D.~Boulanger, M.~Bruynooghe, and M.~Denecker.
\newblock Abstracting s-semantincs using a model-theoretic approach.
\newblock In {\em Proceedings PLILP'94 (Programming Language Implementation and
  Logic programming)}, LNCS 844, pages 432--446. Springer-Verlag, 1994.

\bibitem{IB}
M.~Bruynooghe.
\newblock Solving combinatorial search problems by intelligent backtracking.
\newblock {\em Information Processing Letters}, 12(1):36--39, 1981.

\bibitem{PLILP98}
M.~Bruynooghe, H.~Vandecasteele, D.A. de~Waal, and M.~Denecker.
\newblock Detecting unsolvable queries for definite logic programs.
\newblock In {\em Principles of Declarative Programming, Proc. PLILP'98,
  ALP'98}, LNCS 1490, pages 118--133. Springer-Verlag, 1998.

\bibitem{CharatonikPodelski-SAS98}
W.~Charatonik and A.~Podelski.
\newblock Directional type inference for logic programs.
\newblock In Giorgio Levi, editor, {\em Proceedings of the Fifth International
  Static Analysis Symposium (SAS)}, LNCS 1503, pages 278--294, Pisa, Italy,
  1998. Springer-Verlag.

\bibitem{SAS94}
M.~Codish and B.~Demoen.
\newblock Deriving polymorphic type dependencies for logic programs using
  multiple incarnations of prop.
\newblock In {\em Proceedings SAS'94 (Static Analysis Symposium)}, LNCS 864,
  pages 281--296. Springer-Verlag, 1994.

\bibitem{CD95}
M.~Codish and B.~Demoen.
\newblock Analysing logic programs using ``prop''-ositional logic programs and
  a magic wand.
\newblock {\em Journal of Logic Programming}, 25(3):249--274, December 1995.
\newblock First version in ILPS'93.

\bibitem{survey}
D.~{De Schreye} and S.~Decorte.
\newblock Termination of logic programs: the never-ending story.
\newblock {\em Journal of Logic Programming}, 19 \& 20:199--260, May/July 1994.

\bibitem{CPD}
D.~De~Schreye, R.~Gl\"{u}ck, J.~J{\o}rgensen, M.~Leuschel, B.~Martens, and M.H.
  S{\o}rensen.
\newblock Conjuctive partial deduction: foundations, control, algorithms and
  experiments.
\newblock {\em J. Logic Programming}, 1999.
\newblock To appear.

\bibitem{IJCAI97}
D.A. de~Waal, M.~Denecker, M.~Bruynooghe, and M.~Thielscher.
\newblock The generation of pre-interpretations for detecting unsolvable
  planning problems.
\newblock In U.~Furbach, editor, {\em Proc. IJCAI Workshop on Model based
  Automated reasoning}, pages 103--112, 1997.

\bibitem{SLDNFA}
M.~Denecker and D.~De Schreye.
\newblock {SLDNFA}: an abductive procedure for abductive logic programs.
\newblock {\em Journal of Logic Programming}, 34(2):201--226, Februari 1998.

\bibitem{Enderton72}
H.~B. Enderton.
\newblock {\em A Mathematical Introduction To Logic}.
\newblock Academic Press, 1972.

\bibitem{Gal}
J.~Gallagher, D.~Boulanger, and H.~Sa\v{g}lam.
\newblock Practical model-based static analysis for definite logic programs.
\newblock In J.~Lloyd, editor, {\em Proc. ILPS'95}, pages 351--365, Portland,
  Oregon, December 1995. MIT Press.

\bibitem{Gal94}
J.P. Gallagher and D.A. de~Waal.
\newblock Fast and precise regular approximations of logic programs,.
\newblock In P.~Van~Hentenryck, editor, {\em Proc. ICLP94}, pages 599--613. MIT
  Press, 1994.

\bibitem{Heintze-sc-anal-92}
N.~Heintze.
\newblock {\em Set based program analysis}.
\newblock PhD thesis, School of Computer Science, Carnegie Mellon University,
  1992.

\bibitem{Kakas}
A.C. Kakas, R.A. Kowalski, and F.~Toni.
\newblock Abductive logic programming.
\newblock {\em J. Logic and Computation}, 2(6):719--770, 1992.

\bibitem{LeuschelDeSchreyeDeWaal:JICSLP96}
M.~Leuschel, D.~De~Schreye, and D.A. de~Waal.
\newblock A conceptual embedding of folding into partial deduction: Towards a
  maximal integration.
\newblock In Michael Maher, editor, {\em Proceedings of the Joint International
  Conference and Symposium on Logic Programming JICSLP'96}, pages 319--332,
  Bonn, Germany, September 1996. MIT Press.

\bibitem{Mishra-84}
P.~Mishra.
\newblock Towards a theory of types in {Prolog}.
\newblock In {\em IEEE International Symposium on Logic Programming}, pages
  289--298, 1984.

\bibitem{Peltier98}
N.~Peltier.
\newblock A new method for automated finite model building exploiting failures
  and symmetries.
\newblock {\em J. Logic and Computation}, 8(4):511--543, 1998.

\bibitem{XSB:ACM}
K.~Sagonas, T.~Swift, and D.S. Warren.
\newblock {XSB} as an efficient deductive database engine.
\newblock In {\em Proceedings of SIGMOD 1994 Conf. ACM}. Acm Press, 1994.

\bibitem{ydshen}
Y.D. Shen.
\newblock An extended variant of atoms loop check for positive logic programs.
\newblock {\em New Generation Computing}, 15(2):187--203, 1997.

\bibitem{FINDER1}
J.~Slaney.
\newblock {FINDER}: Finite domain enumerator system description.
\newblock Technical Report TR-ARP-2-94, Centre for Information Science
  Research, Australian National University, Australia, 1994.
\newblock Also in Proc. CADE-12.

\bibitem{FINDER}
J.~Slaney.
\newblock {FINDER} - finite domain enumerator - version 3.0 - notes and guide.
\newblock Technical report, Centre for Information Science Research, Australian
  National University, Australia, 1997.

\bibitem{TPTP}
C.B. Suttner and G.~Sutcliffe.
\newblock The {TPTP} problem library, version 2.1.0, 15/12/1997.
\newblock Technical Report 97/08, James Cook University, Australia, 1997.

\bibitem{SICSTUS}
Swedish Institute of Computer Science.
\newblock {\em SICSTUS Prolog User's Manual}, November 1997.

\bibitem{OLDT}
H.~Tamaki and T.~Sato.
\newblock {OLD resolution with tabulation}.
\newblock In Ehud Shapiro, editor, {\em Proceedings ICLP'86}, Lecture Notes in
  Computer Science, pages 84--98, London, 1986. Springer-Verlag.

\bibitem{hent89}
P.~Van~Hentenryck.
\newblock {\em Constraint Satisfaction in Logic Programming}.
\newblock The MIT press, 1989.

\bibitem{Memoing@CACM-92}
D.S. Warren.
\newblock {Memoing for Logic Programs}.
\newblock {\em Communications of the ACM}, 35(3):93--111, March 1992.

\bibitem{Weidenbach97jar}
C.~Weidenbach.
\newblock {SPASS} version 0.49.
\newblock {\em Journal of Automated Reasoning}, 18(2):247--252, 1997.

\bibitem{SEM}
J.~Zhang and H.~Zhang.
\newblock {SEM}: a system for enumerating models.
\newblock In {\em Proc. IJCAI-95, {\em Vol.\ 1}}, pages 298--303. Morgan
  Kaufmann, 1995.

\end{thebibliography}

\end{small}

\section*{Appendix}
\label{sec:app}

\subsection{Prolog code of examples}
\label{sec:prolog}

\paragraph{odd\_even.}
Here and further on, the considered queries are
the bodies of 0-arity predicates. 
\begin{verbatim}
even(zero).  
even(s(X)) :- odd(X).  
odd(s(X)) :- even(X).  
odd_even :- even(X), even(s(X)).  
\end{verbatim}

\paragraph{wicked\_oe.}
A version of {\tt odd\_even} with an extra superfluous argument which creates
a term with 4 different functors.

\begin{verbatim}
wicked_even(zero,U) :- wicked_p(U).  
wicked_even(s(X),U) :- wicked_odd(X,_V), wicked_p(U).  
wicked_odd(s(X),U) :- wicked_even(X,_V), wicked_p(U).
wicked_p(f(g(h(a)))).  
wicked_oe :- wicked_even(X, _U1), wicked_even(s(X),_U2).
\end{verbatim}

\paragraph{appendlast.}
Appending $[a]$ to a list cannot yield a list ending in a $b$. 

\begin{verbatim}
app([],L,L).
app([H|X],Y,[H|Z]) :- app(X,Y,Z).
last([X],X).
last([H,H2|T],X) :- last([H2|T],X).
appendlast:- app(X, [a], Xs), last(Xs, b).
\end{verbatim}

\paragraph{reverselast.}
If the accumulator is initialised with a list ending in $a$, then the
result of the call to reverse cannot be a list ending in a $b$.

\begin{verbatim}
last([X],X).
last([H,H2|T],X) :- last([H2|T],X).
reva([], Acc, Acc).
reva([Y|Z], R, Acc):- reva(Z, R, [Y|Acc]).
reverselast:- reva(L, R, [a]), last(R, b).
\end{verbatim}

\paragraph{nreverselast.}
Reversing a list beginning with an $a$ cannot result in a list ending
in $b$.

\begin{verbatim}
rev([], []).
rev([X|Y], R):- rev(Y, S), app(S, [X], R).
app([],L,L).
app([H|X],Y,[H|Z]) :- app(X,Y,Z).
last([X],X).
last([H,H2|T],X) :- last([H2|T],X).
nreverselast :- rev([a|X], R), last(R, b).
\end{verbatim}

\paragraph{schedule.}
There cannot be 2 $c$'s in a list with all $n$'s but the first
element.

\begin{verbatim}
mv(R):- tr(R,NewR), mv(NewR).
mv(R):- atleast2c(R).   % success iff R is non-safe state
tr([c,n|Rs], [n,c|Rs]).    
tr([n|Rs], [n|NewRs]):- tr(Rs,NewRs).
tr([],[]).
cFirst([c|Qs]):- nOnly(Qs).
nOnly([n|Qs]):- nOnly(Qs).
nOnly([n]).
atleast2c([c|L]):- atleast1c(L).
atleast2c([n|L]):- atleast2c(L).
atleast1c([c|_]).
atleast1c([n|L]):- atleast1c(L).
schedule:- cFirst(R), mv(R).
\end{verbatim}

\paragraph{multiseto.}
A program to check that two multisets contain the same elements. The
multiset is represented with a functor {\tt o/2} and a constant {\tt
  emptyMultiSet}. 
This is more a specification than a program. It needs iterative
deepening to find answers.

\begin{verbatim}
sameMultiSet(X, X).  
sameMultiSet(o(X, Y), o(X, Z)) :- sameMultiSet(Y,Z).  
sameMultiSet(o(o(X, Y), Z), U) :- sameMultiSet(o(X, o(Y, Z)), U).
sameMultiSet(U, o(o(X, Y), Z)) :- sameMultiSet(U, o(X, o(Y, Z))).
sameMultiSet(o(emptyMultiSet, X), Y) :- sameMultiSet(X, Y).
sameMultiSet(X, o(emptyMultiSet, Y)) :- sameMultiSet(X, Y).
sameMultiSet(o(X, Y), Z) :- sameMultiSet(o(Y, X), Z).  
multiseto :- sameMultiSet(o(a,o(a,emptyMultiSet)),
                          o(_X,o(emptyMultiSet,b))).
\end{verbatim} 
                         
\paragraph{multisetl.}
The same problem as a normal Prolog program using lists. The query
{\tt ml1} corresponds to the query {\tt m1}. It does not terminate due
to the presence of the variable.
\begin{verbatim} 
sml([], []).  
sml([X|Y], D) :- delete(X, D, E), sml(Y, E).  
delete(M,[M|T], T).  
delete(M, [H|T], [H|L]) :- delete(M, T, L).  
multisetl :- sml([a], X), sml(X, [b]).
\end{verbatim}

\paragraph{blockpair2o.}
A number of planning problems based on a planner due to Michael
Thielsher which operates on a multiset of resources. It has to be
executed under iterative deepening to find plans for most problems
(which have a solution).

In the first problem, the argument collecting the plan ---which is
irrelevant for the existence of a solution--- is omitted. The
multisets are represented using a constant for the empty set and a
binary operator.
\begin{verbatim} 
causesPair(I1, I2):- sameMultiSet(I1, I2).
causesPair(I, G):-  actionPair(C, E), sameMultiSet(o(C, Z), I),
        causesPair(o(E, Z), G).
actionPair(ho(V), o(ta(V),o(cl(V),em))).
actionPair(o(cl(V),o(ta(V),em)), ho(V)).
actionPair(o(ho(V),cl(W)), o(on(V,W),o(cl(V),em))).
actionPair(o(cl(V),o(on(V,W),em)), o(ho(V),cl(W))).
actionPair(o(on(V,W),o(cl(V),em)),
           o(on(s(s(V)),s(V)), o(on(s(V),V), o(on(V,W), 
           o(cl(s(s(V))),em))))).
actionPair(o(on(s(s(V)),s(V)), o(on(s(V),V), o(on(V,W), 
           o(cl(s(s(V))),em)))),
           o(on(V,W), o(cl(V),em))).
sameMultiSet(X, X).
sameMultiSet(o(X, Y), o(X, Z)):- sameMultiSet(Y, Z).
sameMultiSet(o(o(X, Y), Z), U):- sameMultiSet(o(X, o(Y, Z)), U).
sameMultiSet(U, o(o(X, Y), Z)):- sameMultiSet(U, o(X, o(Y, Z))).
sameMultiSet(o(emptyMultiSet, X), Y):- sameMultiSet(X, Y).
sameMultiSet(X, o(emptyMultiSet, Y)):- sameMultiSet(X, Y).
sameMultiSet(o(X, Y), Z):- sameMultiSet(o(Y, X), Z).
blockpair2o:-
     causesPair(o(on(s(nul),nul), o(ta(nul), o(cl(s(nul)),em))),
            o(on(s(s(nul)),s(nul)), o(on(s(nul),nul), o(ta(nul),
                                        o(cl(s(s(nul))),em))))).
\end{verbatim}

\paragraph{blockpair3o.}
Same problem but with the extra argument to collect the plan.

\begin{verbatim}
causesPair(I1, void, I2):- sameMultiSet(I1, I2).
causesPair(I, plan(A, P), G):- actionPair(C, A, E), 
        sameMultiSet(o(C, Z), I), causesPair(o(E, Z), P, G).

actionPair(ho(V),put_down(V),o(ta(V),o(cl(V),em))).                           
actionPair(o(cl(V),o(ta(V),em)),pick_up(V),ho(V)).                            
actionPair(o(ho(V),cl(W)),stack(V,W),o(on(V,W),o(cl(V),em))).
actionPair(o(cl(V),o(on(V,W),em)),unstack(V),o(ho(V),cl(W))).
actionPair(o(on(V,W),o(cl(V),em)),add_two,                                     
           o(on(s(s(V)),s(V)), o(on(s(V),V), 
             o(on(V,W), o(cl(s(s(V))),em))))).
actionPair(o(on(s(s(V)),s(V)), o(on(s(V),V), 
             o(on(V,W), o(cl(s(s(V))),em)))),
           delete_two,
           o(on(V,W), o(cl(V),em))).
sameMultiSet(X, X).
sameMultiSet(o(X, Y), o(X, Z)):- sameMultiSet(Y, Z).
sameMultiSet(o(o(X, Y), Z), U):- sameMultiSet(o(X, o(Y, Z)), U).
sameMultiSet(U, o(o(X, Y), Z)):- sameMultiSet(U, o(X, o(Y, Z))).
sameMultiSet(o(emptyMultiSet, X), Y):- sameMultiSet(X, Y).
sameMultiSet(X, o(emptyMultiSet, Y)):- sameMultiSet(X, Y).
sameMultiSet(o(X, Y), Z):- sameMultiSet(o(Y, X), Z).
blockpair3o:-
     causesPair(o(on(s(nul),nul), o(ta(nul), o(cl(s(nul)),em))), 
                _Plan,
                o(on(s(s(nul)),s(nul)),o(on(s(nul),nul),o(ta(nul),
                                          o(cl(s(s(nul))),em))))).
\end{verbatim}

\paragraph{blockpair2l.}
The next planner represents resources as a list. No argument to
collect the plan.

\begin{verbatim}
causesPairl(I,I).  
causesPairl(I,G) :- actionPairl(C,E), m_subset(C,I,Z), app(E,Z,S), 
                    causesPairl(S,G).
actionPairl([ho(V)],[ta(V),cl(V),em]).  
actionPairl([cl(V),ta(V),em], [ho(V)]).
actionPairl([ho(V),cl(W)], [on(V,W),cl(V),em]).
actionPairl([cl(V),on(V,W),em], [ho(V),cl(W)]).
actionPairl([on(V,W),cl(V),em],
            [on(s(s(V)),s(V)),on(s(V),V),on(V,W),cl(s(s(V))),em]).
actionPairl([on(s(s(V)),s(V)),on(s(V),V),on(V,W),cl(s(s(V))),em],
            [on(V,W),cl(V),em]).
m_subset([], L, L).
m_subset([H|T], L1, L2):-  delete(H, L1, L3), m_subset(T, L3, L2).
delete(M, [M|T], T).
delete(M, [H|T], [H|L]):-  delete(M, T, L).
app([], X, X).
app([X|Y], Z, [X|W]):-  app(Y, Z, W).
blockpair2l :- 
      causesPairl([on(s(0),0),ta(0),cl(s(0)),em], Sequence),
      m_subset([on(s(s(0)),s(0)),on(s(0),0),ta(0),
                         cl(s(s(0))),em], Sequence, []).
\end{verbatim}

\paragraph{blockpair3l.}
With an extra argument to collect the plan.

\begin{verbatim}
causesPairl(I,void,I).                   
causesPairl(I,plan(A,P),G) :- actionPairl(C,A,E), m_subset(C,I,Z),
        app(E,Z,S), causesPairl(S,P,G).
actionPairl([ho(V)],put_down(V),[ta(V),cl(V),em]).                            
actionPairl([cl(V),ta(V),em],pick_up(V),[ho(V)]).                             
actionPairl([ho(V),cl(W)],stack(V,W),[on(V,W),cl(V),em]).                     
actionPairl([cl(V),on(V,W),em],unstack(V),[ho(V),cl(W)]).                     
actionPairl([on(V,W),cl(V),em],add_two,                                     
            [on(s(s(V)),s(V)),on(s(V),V),on(V,W),cl(s(s(V))),em]).             
actionPairl([on(s(s(V)),s(V)),on(s(V),V),on(V,W),cl(s(s(V))),em],
            delete_two, [on(V,W),cl(V),em]).                                             
m_subset([], L, L).
m_subset([H|T], L1, L2):-  delete(H, L1, L3), m_subset(T, L3, L2).
delete(M, [M|T], T).
delete(M, [H|T], [H|L]):-  delete(M, T, L).
app([], X, X).
app([X|Y], Z, [X|W]):-  app(Y, Z, W).
blockpairl:-
        causesPairl([on(s(nul),nul),ta(nul),cl(s(nul)),em], 
                    _Plan, Sequence),
        m_subset([on(s(s(nul)),s(nul)),on(s(nul),nul),
                  ta(nul),cl(s(s(nul))),em], Sequence, []).
\end{verbatim}

\paragraph{blocksol.}
Finally a case where there exists a solution.

\begin{verbatim}
actionZerol([ho(V)], [ta(V), cl(V), em]).  
actionZerol([cl(V), ta(V),em], [ho(V)]).  
actionZerol([ho(V), cl(W)], [on(V,W), cl(V), em]).
actionZerol([cl(V), on(V, W), em], [ho(V), cl(W)]).
actionZerol([on(X, Y), cl(X), em], 
            [on(s(X), X), on(X, Y), cl(s(X)),em]).
causesZerol(I, I).  
causesZerol(I, G) :- actionZerol(C, E),  m_subset(C,I, Z), 
                     app(E, Z, S), causesZerol(S, G).
m_subset([], L, L).
m_subset([H|T], L1, L2):-  delete(H, L1, L3), m_subset(T, L3, L2).
delete(M, [M|T], T).
delete(M, [H|T], [H|L]):-  delete(M, T, L).
app([], X, X).
app([X|Y], Z, [X|W]):-  app(Y, Z, W).
blocksol :- causesZerol([on(s(0),0),ta(0),cl(s(0)),em],
            [on(s(s(0)),s(0)),on(s(0),0),ta(0),cl(s(s(0))),em]).  
\end{verbatim}

\subsection{An example of input for FINDER}
\label{sec:inputfinder}
FINDER supports only nulladic, monadic and dyadic functions. As
predicates are declared as functions with {\bf boolean} as value sort,
these restrictions also apply to predicate definitions and forces some
transformation for predicates with arity $> 2$: for example a ternary
atom $p(t_1,t_2,t_3)$ is encoded as $p(t_1,arg_p(t_2,t_3))$ where
$arg_p/2$ is a binary functor used to encode atoms with predicate p/3
and which constructs a term in a sort different from the sort of the
terms $t_1,t_2$, and $t_3$ of the original atom. To have the same
expressivity as the original p based on a 2 element domain
pre-interpretation, the pre-interpretation of $arg_p$ has to be based
on a 4 element domain.

Input for the {\tt multisetl} problem. The ternary predicate {\tt delete/3}
has been converted in a binary {\tt delete\_was3/2} predicate. Besides
the sort {\tt term} for all terms of the original program, a sort {\tt
  pair} has been introduced. The extra functor {\tt delete\_argpair} is
used to bundle two terms of sort {\tt term} into one of sort {\tt
  pair}.

\begin{verbatim}
sort { term cardinality = 2.
       pair cardinality = 4.
     }

const {a: term.
       b: term.
       nil: term.
       ml1: bool.
      }

function {  cons: term, term -> term.
            delete_argpair: term, term -> pair.
            delete_was3: term, pair -> bool.
            sml: term, term -> bool.
         }
clause {ml1 -> false.
        sml(cons(a, nil), z), sml(z, cons(b, nil)) -> ml1. 
        sml(nil, nil).
        delete_was3(y, delete_argpair(w, z)), sml(x, z) 
                  ->  sml(cons(y, x), w). 
        delete_was3(z, delete_argpair(cons(z, y), y)).
        delete_was3(w, delete_argpair(x, z)) 
        -> delete_was3(w, delete_argpair(cons(y, x), cons(y, z))). 
       }
setting {solutions: 1
         verbosity {models: brief   stats: full    job: brief
        }
\end{verbatim}

\end{document}